\documentclass[10pt,final,journal]{IEEEtran}
\normalsize

\ifCLASSINFOpdf
\else
\fi

\usepackage[utf8]{inputenc}
\usepackage[english]{babel}
\usepackage{graphicx}
\usepackage{float}
\usepackage{amssymb}
\usepackage{amsmath}
\usepackage [autostyle, english = american]{csquotes}
\usepackage{mathptmx}
\usepackage{multirow}
\usepackage{times}
\usepackage{calrsfs}
\usepackage{colortbl}
\usepackage{amsthm}
\usepackage{amsfonts}
\usepackage{array,booktabs,tabularx}
\newcolumntype{M}[1]{>{\centering\arraybackslash}m{#1}}
\DeclareMathAlphabet{\pazocal}{OMS}{zplm}{m}{n}
\MakeOuterQuote{"}
\usepackage[bookmarks=false]{hyperref}
\usepackage{caption}
\usepackage{subcaption}
\newtheorem{theorem}{Theorem}
\newtheorem{proposition}[theorem]{Proposition}
\usepackage{xcolor}
\usepackage{etoolbox}
\usepackage[export]{adjustbox}
\usepackage{lipsum}
\makeatletter
\DeclareRobustCommand{\change}{%
  \@bsphack
  \leavevmode
  \color{red}%
  \@esphack
}
\DeclareRobustCommand{\stopchange}{%
  \@bsphack
  \normalcolor
  \@esphack
}
\makeatletter
\@ifclasslater{IEEEtran}{2012/12/26}
 {\newcommand{\killproofname}{\unskip\nopunct}}
 {\newcommand{\killproofname}[1]{\unskip\aftergroup\ignorespaces\ignorespaces}}
\makeatother

\begin{document}
%

\title
{Deep Learning-based Codebook Design for Code-Domain Non-Orthogonal Multiple Access Approaching Single-User Bit-Error Rate Performance}
%
%
%
\author{Minsig Han, Hanchang Seo, Ameha T.~Abebe
       and~Chung G.~Kang,~\IEEEmembership{Senior Member,~IEEE,}

}

%
%
\markboth{IEEE Transactions on Cognitive Communications and Networking}%
{Submitted paper}
%



\maketitle
\begin{abstract}
A general form of codebook design for code-domain non-orthogonal multiple access (CD-NOMA) can be considered equivalent to an autoencoder (AE)-based constellation design for multi-user multidimensional modulation (MU-MDM). Due to a constrained design space for optimal constellation, e.g., fixed resource mapping and equal power allocation to all codebooks, however, existing AE architectures produce constellations with suboptimal bit-error-rate (BER) performance. Accordingly, we propose a new architecture for MU-MDM AE and underlying training methodology for joint optimization of resource mapping and a constellation design with bit-to-symbol mapping, aiming at approaching the BER performance of a single-user MDM (SU-MDM) AE model with the same spectral efficiency. The core design of the proposed AE architecture is dense resource mapping combined with the novel power allocation layer that normalizes the sum of user codebook power across the entire resources. This globalizes the domain of the constellation design by enabling flexible resource mapping and power allocation. Furthermore, it allows the AE-based training to approach a global optimal MU-MDM constellations for CD-NOMA. Extensive BER simulation results demonstrate that the proposed design outperforms the existing CD-NOMA designs while approaching the single-user BER performance achieved by the equivalent SU-MDM AE within 0.3 dB over the additive white Gaussian noise channel.

\end{abstract}


%
\IEEEpeerreviewmaketitle

\section{Introduction}
%
%
%
%
\IEEEPARstart{S}{ince} the early days of 5G research and development, non-orthogonal multiple access (NOMA) schemes have gained considerable attention owing to their superior spectral efficiency, as compared with orthogonal multiple access (OMA) [1]. The allows for sharing time/frequency resources among users by separating them in either power or code domains. The power-domain NOMA (PD-NOMA) utilizes superposition coding at the transmitter and successive interference cancellation at the receiver by exploiting the channel gain difference between multiplexed users [2]. In code-domain NOMA (CD-NOMA), meanwhile, different users are allocated different codes and multiplexed over the same time/frequency resources. Furthermore, variants of the NOMA schemes have been proposed for grant-free access so as to reduce the access delay and signaling overhead [3]-[5]. Because a user symbol is transmitted over multiple complex dimensions (resources) in CD-NOMA, it inherently possesses both the nature of the modulation and channel coding (diversity).

Sparse-code multiple access (SCMA) has been considered as a promising candidate among the notable CD-NOMA schemes, owing to its simplicity and superior performance [6]-[13]. It evenly distributes the number of users that interfere with each other over some resources while exploiting constellation rotations to maximize the Euclidean distance (ED) among different codebooks (CBs). The sparse CBs in SCMA, as compared with the dense CBs, restrict the decoding complexity of the message passing algorithm (MPA) to grow exponentially with only a fraction of the number of superposed users, instead of all the users. In general, all CD-NOMA schemes, including SCMA, achieve a higher spectral efficiency than OMA at the expense of detection complexity. Another CD-NOMA scheme that has gained considerable traction is pattern division multiple access (PDMA) [14]. The CBs for PDMA are designed such that there is a disparity in the transmit diversity order among superposed users (layers). This enables a successive interference cancellation (SIC)-based receiver to decode a user’s signal in the descending diversity order. It is also worth mentioning that power domain (PD)-NOMA induces power disparity among users. Furthermore, PD-NOMA is considered as a special case of CD-NOMA, wherein a user’s symbol is transmitted over a single complex dimension. In general, all of these NOMA schemes offer a clear performance advantage over OMA. However, CD-NOMA takes a complicated optimization process for codeword (CW) design along with high-complexity multi-user detection, making the conventional and practical design schemes far from the optimum performance.

Following the seminal paper with deep learning (DL) by O’shea et al. [15], in particular, an end-to-end optimized architecture that exploits the similarity between a communication system and an autoencoder (AE) has gained considerable attention [16]–[27]. In [15], the AE for the end-to-end learning of a communication system is used to optimize a multi-dimensional modulation (MDM) scheme for a single-user communication. The AE for \textit{single-user} MDM (SU-MDM) in [15], transmitting 4 bits over 7 dimensions, i.e., achieving the spectral efficiency of 4/7 bits/dimension, has been shown to achieve the indistinguishable block error rate (BLER) performance from a conventional system employing a Hamming (7,4) code and maximum likelihood detection (MLD). It is important to note that the AE in [15] was able to design a near-optimal constellation in terms of BLER, provided that the AE training complexity is affordable. Their loss function and hyperparameters of the AE were modified for minimizing the bit-error-rate (BER) in [16]. It improved multi-dimensional constellation shaping by increasing the minimum ED in a constellation with better bit-to-symbol mapping. Furthermore, it is first shown in [27] that an encoder-decoder pair with a near-optimum error performance can be equivalently obtained by perfectly training a special AE, which however might not be possible in practice. Accordingly, [27] proposes a practical AE-based channel coding scheme that is well-suited for receivers with one-bit quantization.

Meanwhile, a superposition of multiuser symbols in CD-NOMA can be treated as one of the signal points from a single constellation of SU-MDM with equivalent spectral efficiency. However, from a multiple-access viewpoint, SU-MDM corresponds to a virtual model for CD-NOMA wherein a user knows the other-users’ simultaneous symbols a priori, i.e., by \textit{Genie} aid. Given the optimized constellation for a single user, it can be extended to the multi-user scenario, in which signal points (codewords) in the multi-dimensional modulation constellation can be divided into multiple subsets, each allocated to the different users, while maintaining the same spectral density as in the single-user scenario. Consequently, the multi-user version of AE which specified different CB for each user can be considered with the same principles as SU-MDM except for the absence of knowledge on the other-users’ symbols. Then in CD-NOMA, as individual constellations must be designed for different users, it is referred to as \textit{multi-user} MDM (MU-MDM). For example, an AE-based SCMA was proposed in [17], generating an improved MU-MDM constellation with sparse resource mapping, while realizing a low-complexity deep neural network (DNN) decoder. It achieves a better BER performance than the conventional SCMA. By contrast, dense code multiple access (DCMA) was proposed in [18] by following the same AE-based design, now for dense resource mapping.

It is important to note that the only difference in the objective of constellation design for SU-MDM and MU-MDM with the same spectral efficiency is that the former tries to maximize the minimum Euclidean distance (ED) of the signal points while the latter tries to maximize the minimum ED of the multi-user superposed signal points. Therefore, when both designs try to transmit the same number of bits through the given signal dimension for the same spectral efficiency, their optimal error performance (i.e., the optimal sphere packing performance with MLD) should be the same. Such optimal error performance can only be achieved if we can perfectly train the corresponding AEs, i.e., fully and globally minimize the loss value [27]. However, such perfect training has not been achieved in the existing literature for MU-MDM design due to inherent and imposed constraints. In the existing MU-MDM constellation designs [17]-[25], the structural constraints in MU-MDM AE design make the performance of MU-MDM AE to be sub-optimal compared to equivalent SU-MDM AE. The structural constraints in MU-MDM AE design as compared to SU-MDM AE design can be summarized into three-fold. The first constraint (referred to as \textbf{Constraint A}) is subject to independent encoder network by input vector split for each user, not necessitating the Genie aid and then, their signal superposition. The second constraint (referred to as \textbf{Constraint B}) is subject to fixed resource mapping, e.g., sparse or dense resource mapping. The last one (referred to as \textbf{Constraint C}) is based on power control constraint with power normalization layer (PNL), which normalizes the user's CB power.

Generally, in solving the optimization problem including constellation optimization, given all the conditions are same, the unconstrained (or constraint-relaxed) optimization problem gives a better or equal performance than the constrained one. Therefore, we want to relax the constraints (equivalently, globalize the search space) for MU-MDM AE structure, aiming at approaching the BER of an equivalent SU-MDM AE design which is not afflicted by any structural constraints of \textbf{Constraint A}, \textbf{Constraint B}, and \textbf{Constraint C}. However, as already described, \textbf{Constraint A} is an inherent (indispensable) constraint in MU-MDM AE architecture. This is because a user’s symbols and the corresponding CWs should be independent of other users’, not necessitating the Genie aid. Thus, the proposed design focuses on relaxing the \textbf{Constraint B} and \textbf{Constraint C} so as to maximize the degree-of-freedom (optimization search space) in AE-based constellation optimization. Accordingly, despite numerous AE-based constellation designs for MU-MDM [17]-[25], we argue that the existing MU-MDM AE structures suffer from unnecessary constraints, e.g., fixed resource mapping and equal power allocation to all CBs, which limits the degrees-of-freedom for the CB design. For the relaxed constraints toward global optimal solution, we propose PNL that normalizes a sum of the user’s CB power across the entire resources. We highlight that, by combining the proposed PNL with dense resource mapping, the MU-MDM AE architecture globalizes the search space to allow AE to readily approach to a global optimal MU-MDM constellation for CD-NOMA. In particular, this combination globalizes the domain of the constellation design by enabling flexible resource mapping and power allocation in multi-user CB design.

Furthermore, the existing AE-based MU-MDM design schemes in [17] and [20]-[21] could not jointly optimize the signal constellation and its bit-to-symbol mapping function. More specifically, they employed L2-norm or cross-entropy with one-hot vector representation as their loss function, trying to minimize the symbol error rate (SER), i.e., without considering a bit-to-symbol mapping function. Moreover, we note that binary vector representation was employed for the loss function to improve the BER performance of the constellation in [18], [22]-[25]. However, the optimal design of the signal points and its bit-to-symbol mapping varies with the signal-to-noise (SNR) level. Therefore, there must be a means to minimize BER over a wide range of SNR levels. Thus, by considering a loss function that simultaneously deals with the ED between signal points and the Hamming distance (HD) between bits assigned to neighboring signal points created through the superposition of multi-user signal points, meaningful hyperparameters can be introduced to adjust the relationship between the ED and HD of the constellation signal points for achieving improved BER performance over a varying SNR range. In this paper, a multi-user generalization of the hyperparameterized loss function and associated two-step training process are proposed to improve a BER performance.

To verify the performance improvement of the proposed MU-MDM design and compare it with equivalent SU-MDM performance, there must be trainable AE architectures for MU-MDM and SU-MDM. In order to handle the complexity of AE architecture as growing exponentially with the number of users, we propose the trainable architectures of deep learning-based designs for MU-MDM and baseline SU-MDM, which can be equivalently compared to each other. They are implemented to demonstrate that the proposed design for MU-MDM can approach the BER performance of baseline SU-MDM design within 0.3 dB in the additive white Gaussian noise (AWGN) channel, thus serving as the best existing design that can be realized with its low-complex decoder.

The rest of the paper is organized as follows. In Section \uppercase\expandafter{\romannumeral2}, we present the system model for CD-NOMA and a problem formulation for its CB design. The limitations of existing studies are discussed thereafter. In Section \uppercase\expandafter{\romannumeral3}, we present a baseline architecture for the AE-based SU-MDM system, which serves as a lower BER performance bound for the proposed MU-MDM scheme subject to the same order of spectral efficiency and complexity. Section \uppercase\expandafter{\romannumeral4} presents the proposed MU-MDM AE structure with different PNL levels and resource utilization. The loss function of the proposed AE is also presented with the underlying training method. Simulation results are presented and discussed in Section \uppercase\expandafter{\romannumeral5}; these results indicate a near single-user BER performance when using the proposed MU-MDM design. Finally, the concluding remarks are collated in Section \uppercase\expandafter{\romannumeral6}.

\section{System Model and Design Approach}
\subsection{System Model for CD-NOMA}
In the CD-NOMA system, $J$-user signals are superposed over $K$ orthogonal resources, where $J>K$ and $J/K$ is referred to as an overloading factor. More specifically, a user’s symbol is mapped to $N$ of the $K$ resources, wherein if $N<K$, the CD-NOMA scheme is said to be sparse; otherwise (i.e., $N=K$), it is dense. Fig. 1 details the bit-to-symbol and symbol-to-resource mapping processes to generate the CWs of multiple users. Assuming $M$-ary symbols, a vector of the transmitted bit sequence for user $j$ is represented as ${{\mathbf{b}}^{(j)}}\in {{\mathbb{B}}^{{{\log }_{2}}M}}$, which is mapped to an $N$-dimensional symbol ${{\mathbf{c}}^{(j)}}\in {{\mathbb{C}}^{N}}$ using a bit-to-symbol mapping function $f_{b}^{(j)}:{{\mathbb{B}}^{{{\log }_{2}}M}}\to {{\mathbb{C}}^{N}}$, $j=1,2,\cdots ,J$. 

For a multi-user modulation scheme that employs a different constellation for different users, let ${{\mathsf{\pazocal{C}}}^{(j)}}$ denote a set of symbols for user $j$, i.e., ${{\mathsf{\pazocal{C}}}^{(j)}}=\{\mathbf{c}_{1}^{(j)},\mathbf{c}_{2}^{(j)},\cdots ,\mathbf{c}_{M}^{(j)}\}$, where $\mathbf{c}_{m}^{(j)}\in {{\mathbb{C}}^{N}}$ and $m=1,2,\cdots ,M$. Furthermore, a constellation for all users is defined as $\mathsf{\pazocal{C}}=\{{{\mathsf{\pazocal{C}}}^{(1)}},{{\mathsf{\pazocal{C}}}^{(2)}},\cdots ,{{\mathsf{\pazocal{C}}}^{(J)}}\}$. In addition, the bit-to-symbol mapping functions for the constellation $\mathsf{\pazocal{C}}$ are represented by the set ${{\mathsf{\pazocal{F}}}_{b}}=\{f_{b}^{(1)},f_{b}^{(2)},\cdots ,f_{b}^{(J)}\}$. Afterward, the $N$-dimensional symbol $\mathbf{c}_{{}}^{(j)}$ for the $j$-th user is mapped to a $K$-dimensional CW ${{\mathbf{s}}^{(j)}}\in {{\mathbb{C}}^{K}}$ using a mapping vector ${{\mathbf{f}}^{(j)}}\in {{\mathbb{B}}^{K}}$, which determines the resources to be selected. More specifically, the $N$ resources for $\mathbf{c}_{{}}^{(j)}$ are selected by the non-zero elements of ${{\mathbf{f}}^{(j)}}\in {{\mathbb{B}}^{K}}$. As an example, consider ${{\mathbf{f}}^{(j)}}={{[1\text{ }0\text{ }1\text{ }0]}^{T}}$ for $K=4$ and $N=2$. This indicates that a two-dimensional complex symbol of the $j$-th user is mapped to the first and third resources. Therefore, a resource mapping matrix for all users is constructed as $\mathbf{F}=[{{\mathbf{f}}^{(1)}},{{\mathbf{f}}^{(2)}},\cdots ,{{\mathbf{f}}^{(J)}}]\in {{\mathbb{B}}^{K\times J}}$. Consequently, each user transmits a CW vector $\mathbf{s}_{{}}^{(j)}$ with its non-zero elements mapped to the resource elements determined by the $j$-th user’s CB, which is denoted as ${{\mathsf{\pazocal{S}}}^{(j)}}=\{\mathbf{s}_{1}^{(j)},\mathbf{s}_{2}^{(j)},\cdots ,\mathbf{s}_{M}^{(j)}\}\in {{\mathbb{C}}^{K}}$. Given $J$, $K$, $N$, and $M$, a CB generation function can be represented as $s\left( \mathsf{\pazocal{C}},{{\mathsf{\pazocal{F}}}_{b}},\mathbf{F};J,K,N,M \right)$, which is an end-to-end function to map the input bit sequences of multiuser to a set of multi-dimension CBs $\mathsf{\pazocal{S}}=\{{{\mathsf{\pazocal{S}}}^{(1)}},{{\mathsf{\pazocal{S}}}^{(2)}},\cdots ,{{\mathsf{\pazocal{S}}}^{(J)}}\}$. Let ${{P}^{(j)}}$ denote the average CB power for the $j$-th user, that is, average power of $M$ CWs in the user $j$’s CB, which is given as ${{P}^{(j)}}=(1/M)\cdot \sum\nolimits_{m=1}^{M}{\left\| \mathbf{s}_{m}^{(j)} \right\|_{2}^{2}}$. 

Now, let us consider a superposition of the CWs from different users, which is represented as $\mathbf{x}\triangleq \sum\nolimits_{j=1}^{J}{{{\mathbf{s}}^{(j)}}}$. It must be noted that $\mathbf{x}$ can take any of ${{M}^{J}}$ realizations, denoted as $\mathsf{\pazocal{X}}$. The received signal is then expressed as $\mathbf{y}=\mathbf{x}+\mathbf{n}$, where $\mathbf{n}\in {{\mathbb{C}}^{K}}$ is a complex AWGN noise vector with $\pazocal{C}\pazocal{N}(0,{{\sigma }^{2}}{{\mathbf{I}}_{K}})$ and $\mathbf{x}\in \mathsf{\pazocal{X}}$. In the receiver, the realization $\mathbf{x}$, i.e., superposed CWs from all the users, can be jointly detected by the following multi-user maximum a posteriori probability detection:
\begin{align}
\mathbf{\hat{x}}=\underset{\mathbf{x}\in \mathsf{\pazocal{X}}}{\mathop{\arg \max }}\,p(\mathbf{x}|\mathbf{y}),
\end{align}
where $p(\mathbf{x}|\mathbf{y})$ is the conditional probability of $\mathbf{x}$, given $\mathbf{y}$. As observed in the system model, the CB of the CD-NOMA system corresponds to the constellation for an MU-MDM system. Therefore, in the following sections, the CWs in a CB are interchangeably referred to as symbols or signal points in a MU-MDM constellation without loss of generality.

\begin{figure}[t]
\centering
\includegraphics[width=\columnwidth,height=1.4in]{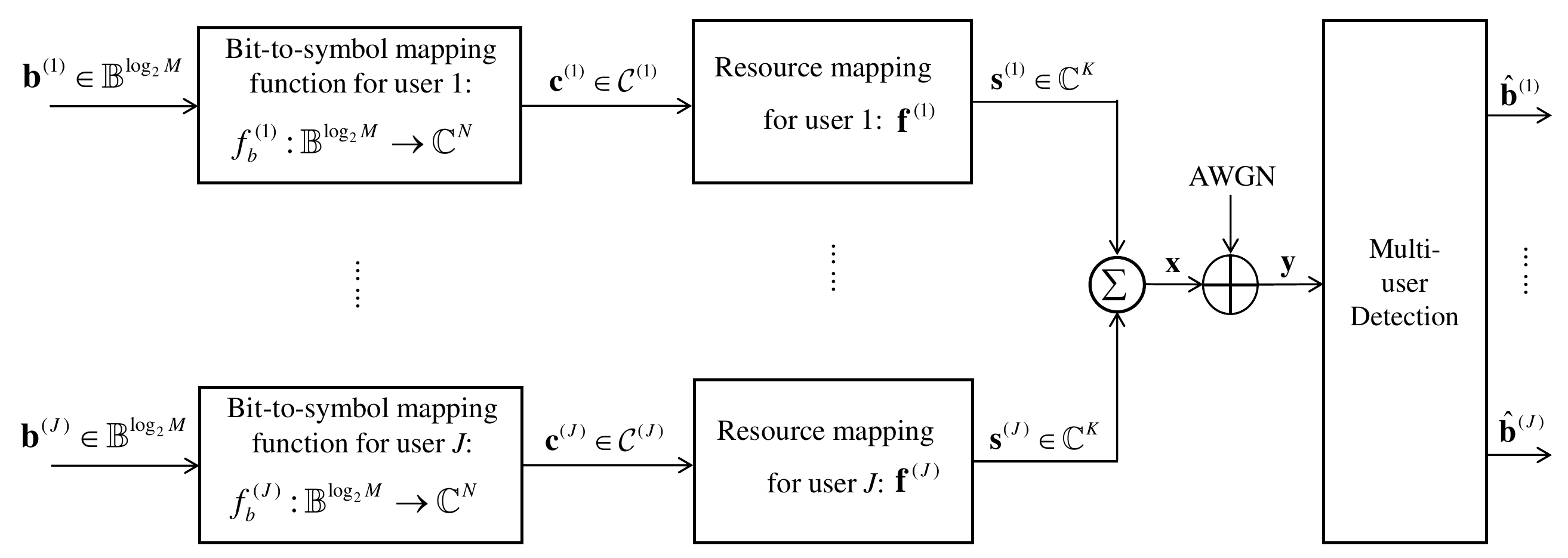}
  \captionsetup{font=small,justification=raggedright, singlelinecheck=false}
\caption{System model for code-domain NOMA with $J$ users and $K$ resources}
\label{fig_sim}
\end{figure}

\subsection{Deep Learning-based MU-MDM Design: Problem Formulation}
In this study, we consider a multi-user CB design problem to determine a constellation set $\mathsf{\pazocal{C}}$, a bit-to-symbol mapping function set ${{\mathsf{\pazocal{F}}}_{b}}$, and a resource mapping matrix $\mathbf{F}$ that optimizes the performance of a multi-user CB generation function $s\left( \mathsf{\pazocal{C}},{{\mathsf{\pazocal{F}}}_{b}},\mathbf{F};J,K,N,M \right)$ for CD-NOMA. It should be noted that the multi-user CB generation $s\left( \mathsf{\pazocal{C}},{{\mathsf{\pazocal{F}}}_{b}},\mathbf{F};J,K,N,M \right)$ corresponds to the MU-MDM. In general, a BER, denoted as the function ${{\varepsilon }_{b}}(\cdot )$, is a typical performance metric for an MU-MDM design. Then, an MU-MDM design problem can be comprehensively formulated through the following optimization problem:
\[({{\mathsf{\pazocal{C}}}^{*}},{{\mathsf{\pazocal{F}}}_{b}}^{*},{{\mathbf{F}}^{*}})=\underset{(\mathsf{\pazocal{C}},{{\mathsf{\pazocal{F}}}_{b}},\mathbf{F})}{\mathop{\arg \min }}\,{{\varepsilon }_{b}}\left( s\left( \mathsf{\pazocal{C}},{{\mathsf{\pazocal{F}}}_{b}},\mathbf{F};J,K,N,M \right) \right) \\ \]
\begin{align}
 & \text{                    s}\text{.t}\text{. }\frac{1}{J}\sum\nolimits_{j=1}^{J}{{{P}^{(j)}}}\le P, 
\end{align}
where $P$ is the total power allocated for the $J$ superposed users. 

Owing to the complexity of (2), the existing approaches attempt to find $\mathsf{\pazocal{C}}$ for the given $\mathbf{F}$, that is, they fail to solve the joint optimal solution for (2). In the existing approaches, specific resource mapping rules are configured when $N$, $J$, $K$, and the detection algorithm, are given. For example, in SCMA, ${{\mathbf{F}}^{*}}$ possesses the property of keeping a minimum number of symbols overlapping the resources in each dimension for the given degree of freedom ${{d}_{f}}=JN/K$ [7]. Furthermore, $N$ is set to be significantly lower than $K$ for sparse $\mathbf{F}$,  to ensure a low complexity MPA that warrants the detection performance. Notably, this complexity for CD-NOMA is given as $O({{M}^{{{d}_{f}}}})$, that is, ${{d}_{f}}<J$ for $N<<K$, which result in lower complexity [6]. Therefore, the optimization problem (2) can be reduced by replacing $\mathbf{F}$ with a handcrafted optimized matrix ${{\mathbf{F}}^{*}}$ while focusing on the constellation signal points and bit-to-symbol mapping optimization alone. For a fixed matrix ${{\mathbf{F}}^{*}}$ and its sparsity level ${{N}^{*}}$, it is restated as
\[({{\mathsf{\pazocal{C}}}^{*}},{{\mathsf{\pazocal{F}}}_{b}}^{*})=\underset{(\mathsf{\pazocal{C}},{{\mathsf{\pazocal{F}}}_{b}})}{\mathop{\arg \min }}\,{{\varepsilon }_{b}}\left( s\left( \mathsf{\pazocal{C}},{{\mathsf{\pazocal{F}}}_{b}},{{\mathbf{F}}^{*}};{{N}^{*}},J,K,M \right) \right)\]
\begin{align}
 & \text{               s}\text{.t}\text{. }\frac{1}{J}\sum\nolimits_{j=1}^{J}{{{P}^{(j)}}}\le P.
\end{align}

However, for DL-based approaches, we argue that the MPA receiver (or any other conventional receiver) can be replaced with a neural network-based multi-user decoder (NN-MUD) receiver, the complexity of which is no longer bounded by the constraint $N$. Thus, it is thus unnecessary to impose any constraint on $N$, as in the existing DL-based approaches, for example, a sparse ${{\mathbf{F}}^{*}}$ with $N<K$ for SCMA [17] or dense ${{\mathbf{F}}^{*}}$ with $N=K$, that is, all the entries of one [18]. Owing to the enormous complexity of $({{M}^{J}})!$ possible combinations for a bit-to-symbol mapping function$f_{b}^{(j)}:{{\mathbb{B}}^{{{\log }_{2}}M}}\to {{\mathbb{C}}^{N}}$, only a signal point in the constellation design was explicitly considered in both the conventional and DL-based approaches [10]–[13], [17]-[18], [20]–[25]. To avoid the complexity of dealing with both $\mathsf{\pazocal{C}}$ and ${{\mathsf{\pazocal{F}}}_{b}}$ for joint optimization, previous studies relied on simple suboptimal approaches, such as a constellation rotation to the mother constellation with a large minimum ED [10]-[13]. Furthermore, there has been no attempt to consider both $\mathsf{\pazocal{C}}$ and ${{\mathsf{\pazocal{F}}}_{b}}$ simultaneously in DL-based MU-MDM design approaches, which would require a new loss function to deal with the multi-user interference for training.

To summarize, the existing MU-MDM design is focused solely on optimizing $\mathsf{\pazocal{C}}$, while neglecting ${{\mathsf{\pazocal{F}}}_{b}}$ under the given constraint of ${{\mathbf{F}}^{*}}$. This implies that there must exist substantial room for reducing the BER, as long as $\mathsf{\pazocal{C}}$, ${{\mathsf{\pazocal{F}}}_{b}}^{{}}$, and $\mathbf{F}$ are jointly optimized to solve (2). Therefore, we considered a more generalized architecture for the MU-MDM AE, along with a more flexible power normalization constraint, which allows for the exploitation of a full degree of freedom for an MU-MDM constellation design. Furthermore, we utilized a hyperparameterized loss function and its corresponding training procedure to optimize ${{\mathsf{\pazocal{F}}}_{b}}^{{}}$ in an explicit manner.

\section{Baseline System: Autoencoder-based SU-MDM}
As discussed earlier, SU-MDM AE can serve as a baseline model for assessing the performance of the proposed MU-MDM design. We note that a generic structure of AE for the design of MDM cannot be immediately extended to SU-MDM, simply because it involves a large size of input vectors. In other words, there must be another architecture that can handle the underlying complexity of SU-MDM, while still providing a performance bound of MU-MDM design with the same spectral efficiency. In this section, we discuss a new trainable architecture for SU-MDM as a baseline model to be compared to MU-MDM. Toward that end, we first present a brief introduction to channel AE, which can represent a CD-NOMA system under design.

\begin{figure}[t]
\centering
\includegraphics[width=0.92\columnwidth,height=1.7in]{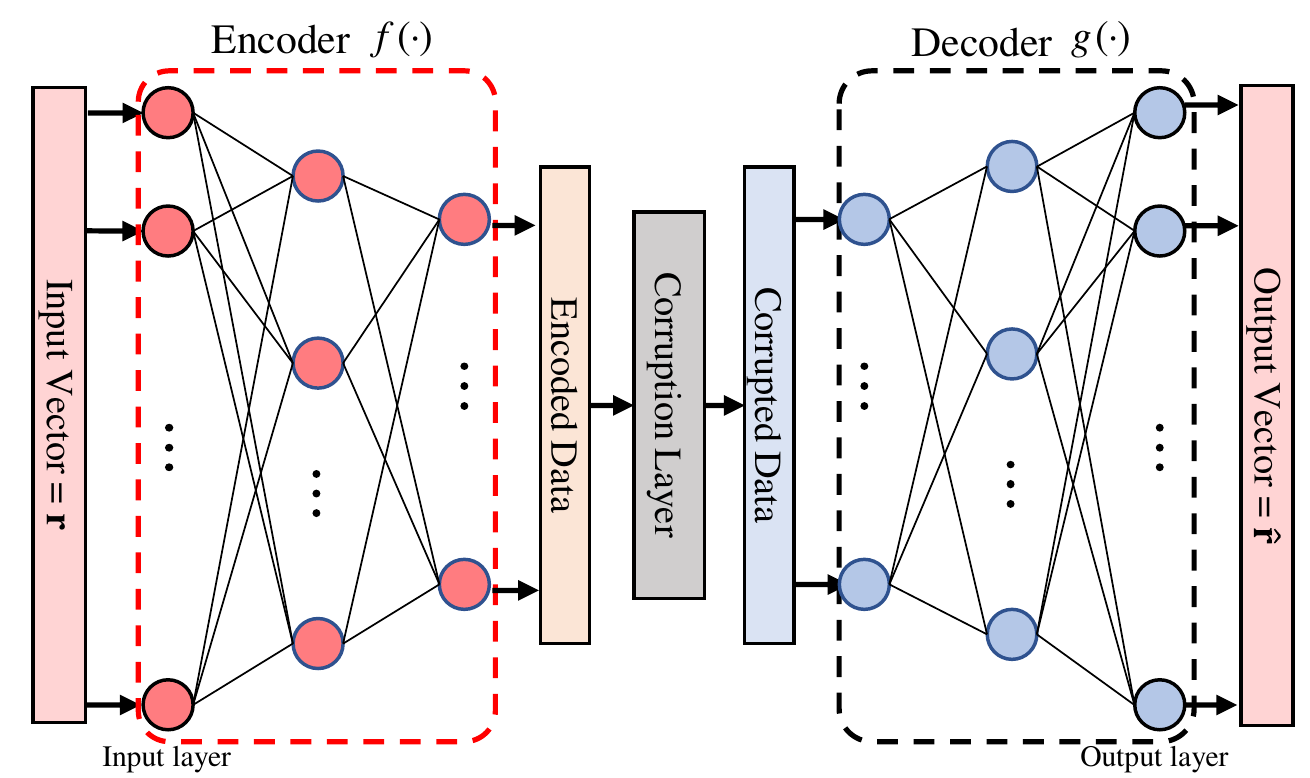}
  \captionsetup{font=small,justification=raggedright, singlelinecheck=false}
\caption{Basic structure of channel AE with corruption}
\label{fig_sim}
\end{figure}

\begin{figure*}[ht]
\centering
\includegraphics[width=0.75\linewidth,height=2.6in]{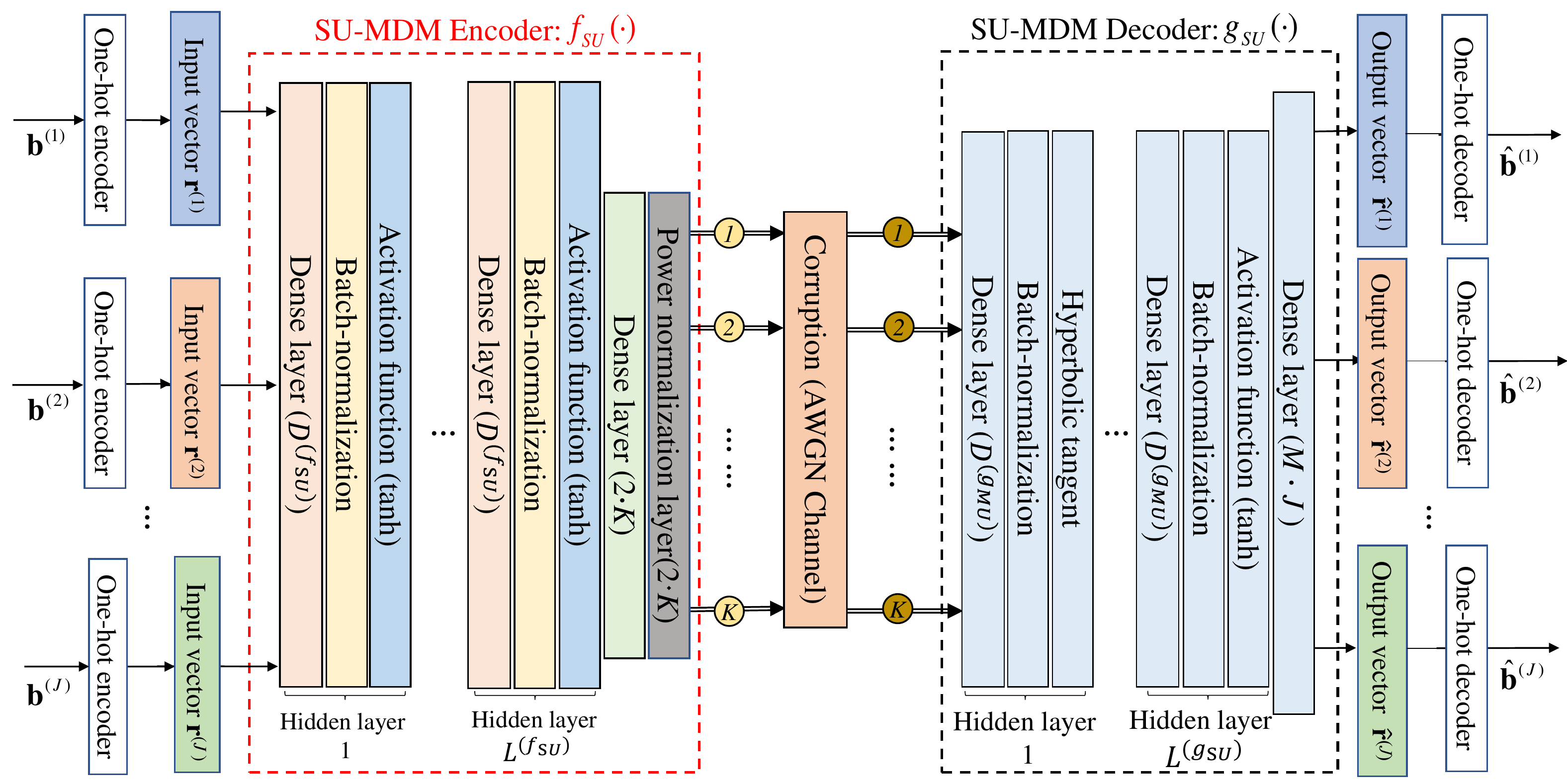}
  \captionsetup{font=small,justification=raggedright, singlelinecheck=false}
\caption{AE structure for SU-MDM model}
\label{fig_sim}
\end{figure*}

\subsection{Channel Autoencoder: Overview}

As shown in Fig. 2, if the channel impairment is applied to the encoded data in a hidden layer, it is termed as a channel AE [15]. The channel AE encodes the input vector $\mathbf{r}$ through the encoder function $f(\cdot )$, and the corruption is subsequently applied to the encoded data. Thereafter, the decoder function $g(\cdot )$ is applied to the corrupted intermediate signal to determine its estimate, denoted as $\mathbf{\hat{r}}$, which needs to be similar (or identical in the perfect sense) to the input vector $\mathbf{r}$, i.e., $\mathbf{\hat{r}}=g(f(\mathbf{r}))\approx \mathbf{r}$. Let ${{\theta }^{(e)}}$ and ${{\theta }^{(d)}}$ be the parameters of the two hidden layer functions, $f(\cdot )$ and $g(\cdot )$, respectively. Given a loss function $L(\cdot )$, the channel AE is trained to minimize it as follows:
\begin{align} 
({{f}_{AE}}^{*},{{g}_{AE}}^{*})=\underset{(f,g)}{\mathop{\arg \min }}\,L\left( \mathbf{r},\mathbf{\hat{r}}\left| {{\theta }^{(f)}},{{\theta }^{(g)}} \right. \right).
\end{align}
The loss function is minimized, as in (4), with an appropriate NN model and hyperparameters. Despite of the corruption in the encoded data, an AE reconstructs the vector $\mathbf{r}$, i.e., $\mathbf{\hat{r}}=\mathbf{r}$, with a probability that approaches one.

In this study, we consider a channel AE in which corruption is given as AWGN $\mathbf{n}$ on the encoded data. In this case, the reconstructed vector can be represented as 
\begin{align} 
\mathbf{\hat{r}}=g\left( f\left( \mathbf{r};{{\theta }^{(f)}} \right)+\mathbf{n};{{\theta }^{(g)}} \right).
\end{align}                   
As in many other practical AE designs, such as in [17] and [22] we also consider the following L2-norm as a baseline loss function:
\begin{align} 
{{L}_{2}}(\mathbf{r},\mathbf{\hat{r}})={{\left\| \mathbf{r}-\mathbf{\hat{r}} \right\|}_{2}}.
\end{align}         
Then, we establish the following proposition to justify our choice of the baseline L2-norm loss function:

\begin{proposition}
\label{theorem1}
Assuming equally likely transmitted symbols over the AWGN channel, any AE with parameters that are perfectly optimized by a loss function of ${{L}_{2}}(\mathbf{r},\mathbf{\hat{r}})={{\left\| \mathbf{r}-\mathbf{\hat{r}} \right\|}_{2}}$ yields the same encoder-decoder pair as that optimized for the minimum ED detection in conventional communication systems for any input vector that uniquely represents the transmitted bit sequence.
\begin{proof}[\killproofname]
\textit{Proof.} \textnormal{See Appendix A.}
\end{proof}
\end{proposition}

In [27], it has already been proven that a channel AE with one-hot coding at the input layer, a softmax activation function at the output layer, and parameters that are optimized via the cross-entropy function can yield the optimal encoder–decoder pair as the one in the conventional MLD for the linear block code, provided it is trained perfectly. Note that L2-norm and cross-entropy loss functions lead to the minimum Euclidean distance (MED)-based and maximum likelihood (ML)-based detection, respectively. Following (A-5), ML-based detection and the MED-based detection achieve the same SERs; thus, both cross-entropy and L2-norm loss functions can interchangeably achieve an optimal encoder-decoder pair that minimizes SER, as in many other existing practical AWGN-based AE architectures [17]-[18], [20]-[25]. Proposition 1 implies that the input vector r does not necessarily need to be a long one-hot input vector when it can be reduced to represent the same transmitted sequence uniquely. In fact, it allows for constructing the equivalent AE with a reduced order of training complexity, especially for SU-MDM, which will serve as a baseline to design the MU-MDM with the same order of training complexity subject to the same bandwidth efficiency.

\begin{figure*}[t]
\raggedright
\begin{subfigure}[b]{0.68\linewidth}
\raggedright
\includegraphics[width=\columnwidth,height=3in]{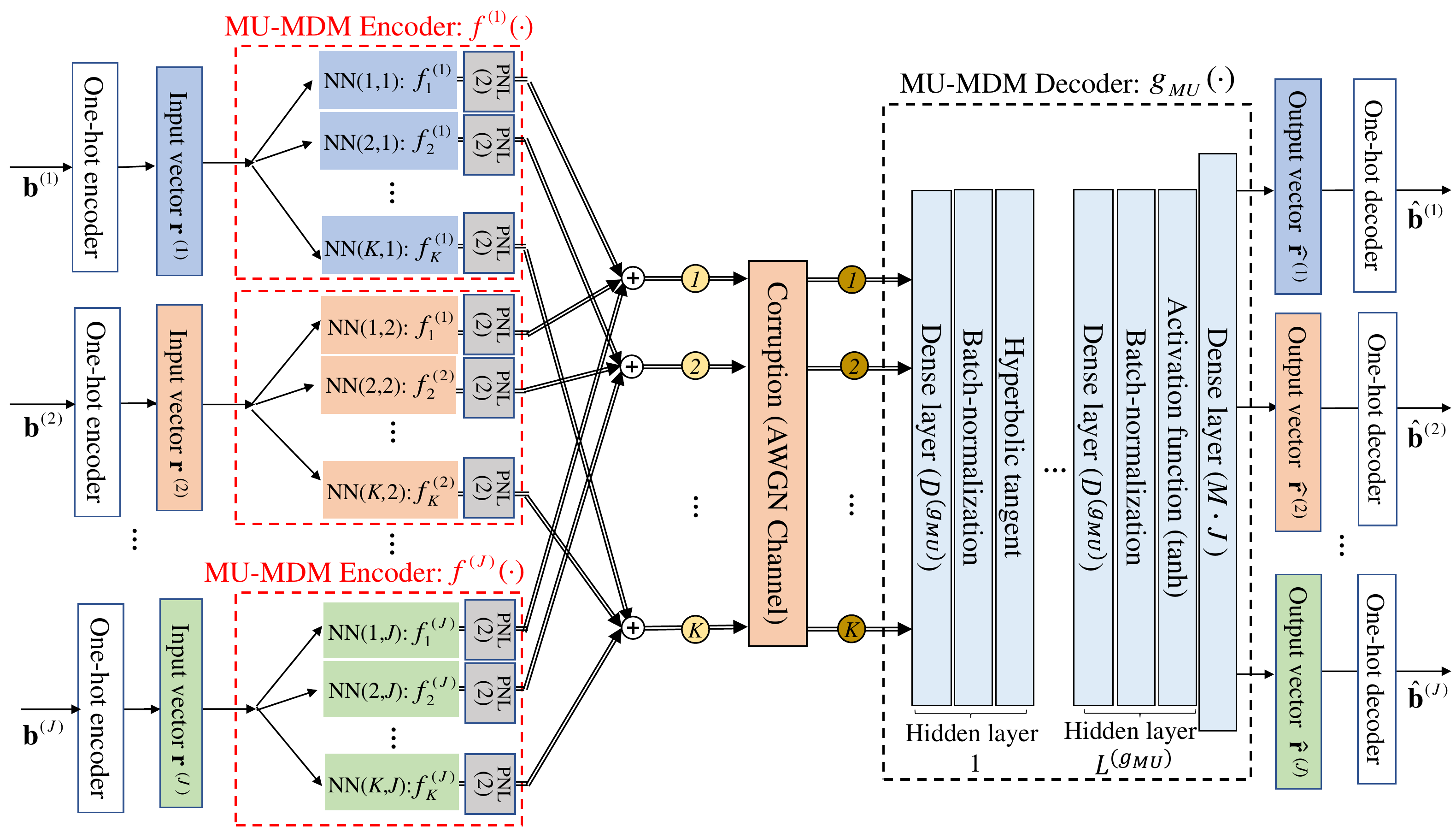}
\label{fig_sim}
\captionsetup{font=small,justification=centering}
\caption{End-to-end configuration}
\end{subfigure}
\raggedright
\begin{subfigure}[b]{0.25\linewidth}
\raggedright
\includegraphics[width=2.4in,height=1.65in]{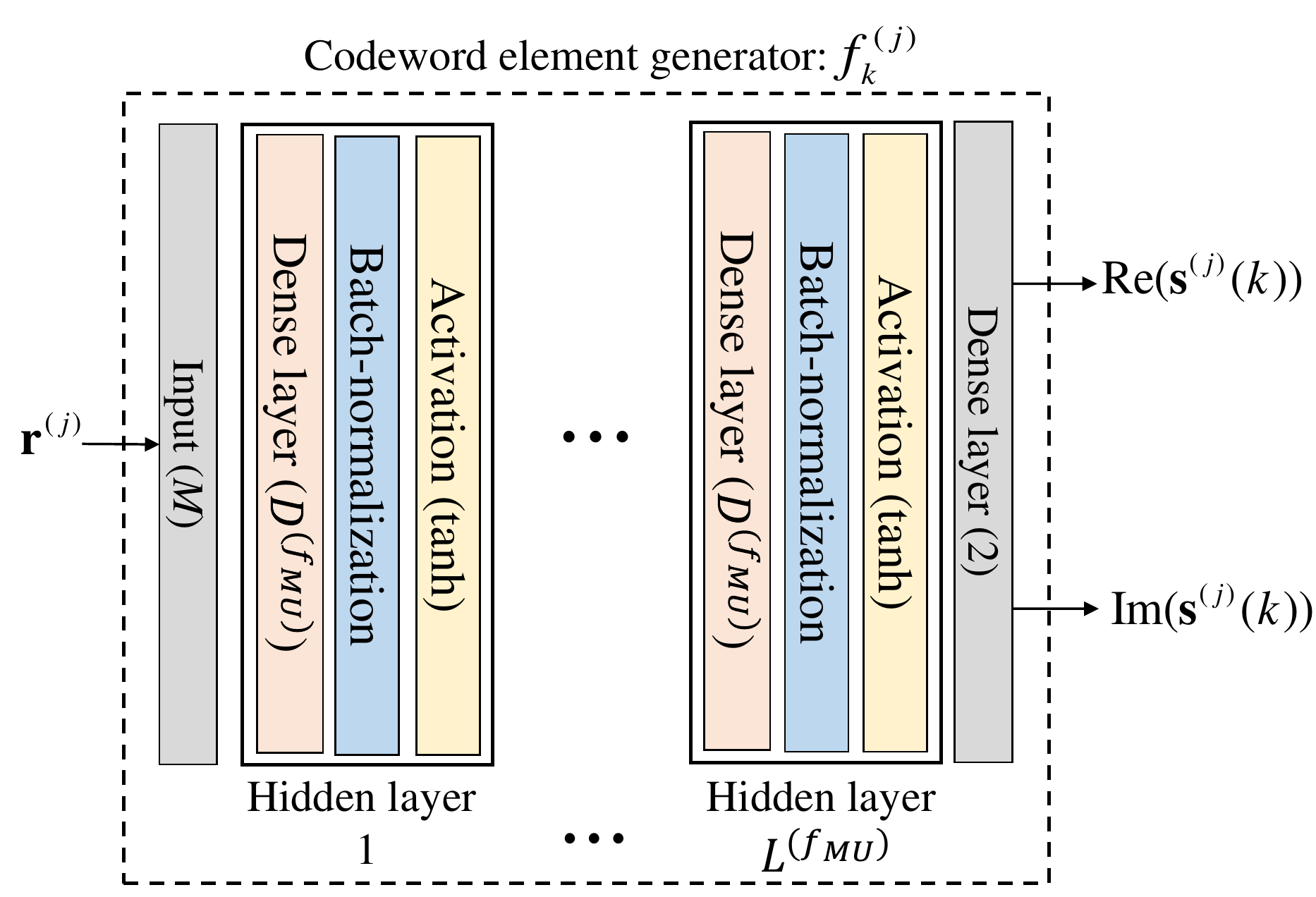}
\label{fig_sim}
\captionsetup{font=small,justification=centering}
\caption{CW element generator for the $j$-th user} 
\end{subfigure}
\captionsetup{font=small,justification=raggedright, singlelinecheck=false}
\caption{The AE structure for MU-MDM model with CW element-wise power normalization}
\end{figure*}

\subsection{AE-based SU-MDM}
In this section, we consider a SU-MDM model with the same bandwidth efficiency as MU-MDM, which transmits $J{{\log }_{2}}M$ bits in $K$-dimensional resources, that is, it achieves the spectral efficiency of $J{{\log }_{2}}M/K$ bits/s/Hz/resource. It is considered to serve as a baseline system for comparison with the proposed MU-MDM system subject to the same bandwidth efficiency. In the existing AE structure for SU-MDM (for example, in [15] and [16]), the input vector $\mathbf{b}$ is encoded to a one-hot vector with a dimension of ${{M}^{J}}$ via a one-hot encoding function $h:{{\mathbb{B}}^{{{\log }_{2}}{{M}^{J}}}}\to {{\mathbb{B}}^{{{M}^{J}}}}$. Let ${{\mathbf{r}}_{i}}\in {{\mathbb{B}}^{{{M}^{J}}}}$ denote one of the one-hot vectors used to represent each input vector $\mathbf{b}$, which is defined as the $i$-th column of the identity matrix ${{\mathbf{I}}_{{{M}^{J}}}}$, that is,
\begin{align}
{{\mathbf{r}}_{i}}={{[\underbrace{0\cdots 0}_{(i-1)\text{ 0 }\!\!'\!\!\text{ s}}\text{ }1\underbrace{0\cdots 0}_{({{M}^{J}}-i)\text{ 0 }\!\!'\!\!\text{ s}}]}^{T}}.
\end{align}
In other words, $\{{{\mathbf{r}}_{i}}\}$ can be used to represent all J symbols uniquely. 

It should be noted that the dimension of the aforementioned one-hot vector can simply explode with $J$, making the NN excessively complex to converge. To reduce the input dimension, multiple one-hot vectors, one for each symbol, can be employed to represent the input vector $\mathbf{b}$. More specifically, consider $J$ independent one-hot vectors, each with a dimension of $M$, which can be concatenated to represent an ${{M}^{J}}$-dimensional one-hot vector. Let ${{\mathbf{r}}^{(j)}}\in {{\mathbb{B}}^{M}}$ denote a one-hot vector to indicate the $j$-th symbol, which is defined as ${{\mathbf{r}}^{(j)}}\in \left\{ \mathbf{r}_{1}^{(j)},\mathbf{r}_{2}^{(j)},\cdots ,\mathbf{r}_{M}^{(j)} \right\}$, where $\mathbf{r}_{i}^{(j)}$ is given by the $i$-th column of the identity matrix ${{\mathbf{I}}_{M}}$, that is, 
\begin{align}
\mathbf{r}_{i}^{(j)}={{[\underbrace{0\cdots 0}_{(i-1)\text{ 0 }\!\!'\!\!\text{ s}}\text{ }1\underbrace{0\cdots 0}_{(M-i)\text{ 0 }\!\!'\!\!\text{ s}}]}^{T}}.
\end{align}
The low-dimensional one-hot vectors are then concatenated as $\mathbf{r}={{[{{({{\mathbf{r}}^{(1)}})}^{T}},{{({{\mathbf{r}}^{(2)}})}^{T}},\cdots ,{{({{\mathbf{r}}^{(J)}})}^{T}}]}^{T}}$, which can be uniquely mapped to a one-hot vector in (7). In fact, ${{M}^{J}}$ signal points in $2K$ dimensions, that is, real and imaginary components of $K$ complex dimensions, are uniquely matched to ${{M}^{J}}$ realizations of the bit sequences by the concatenated one-hot vectors with a total length of $MJ$. Even if the input dimension has been reduced by (8), its performance can still be maintained by Proposition 1. 

Fig. 3 illustrates the overall architecture with the reduced input vectors, which is used as a baseline system for comparison with the proposed MU-MDM system in the next section. The numbers of nodes and layers on the encoder sides are denoted as ${{D}^{({{f}_{SU}})}}$ and ${{L}^{({{f}_{SU}})}}$, respectively. Similarly, the numbers of nodes and layers on the decoder sides are denoted as ${{D}^{({{g}_{SU}})}}$ and ${{L}^{({{g}_{SU}})}}$, respectively.

\begin{figure*}[t]
\centering
\begin{subfigure}[b]{0.28\linewidth}
\centering
\includegraphics[height=2.5in]{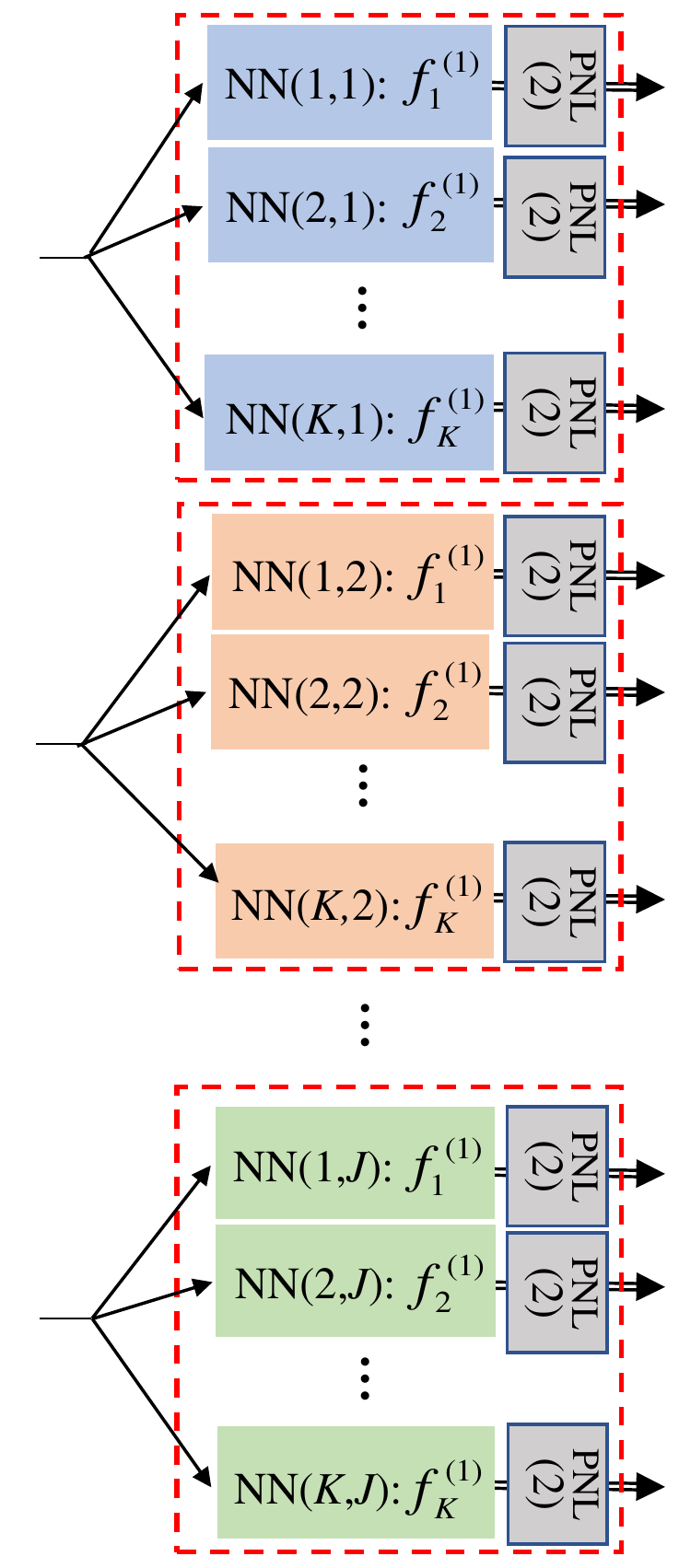}
\label{fig_sim}
\captionsetup{font=small,justification=centering}
  \caption{Level-1 PNL} 
\end{subfigure}
\begin{subfigure}[b]{0.28\linewidth}
\centering
\includegraphics[height=2.5in]{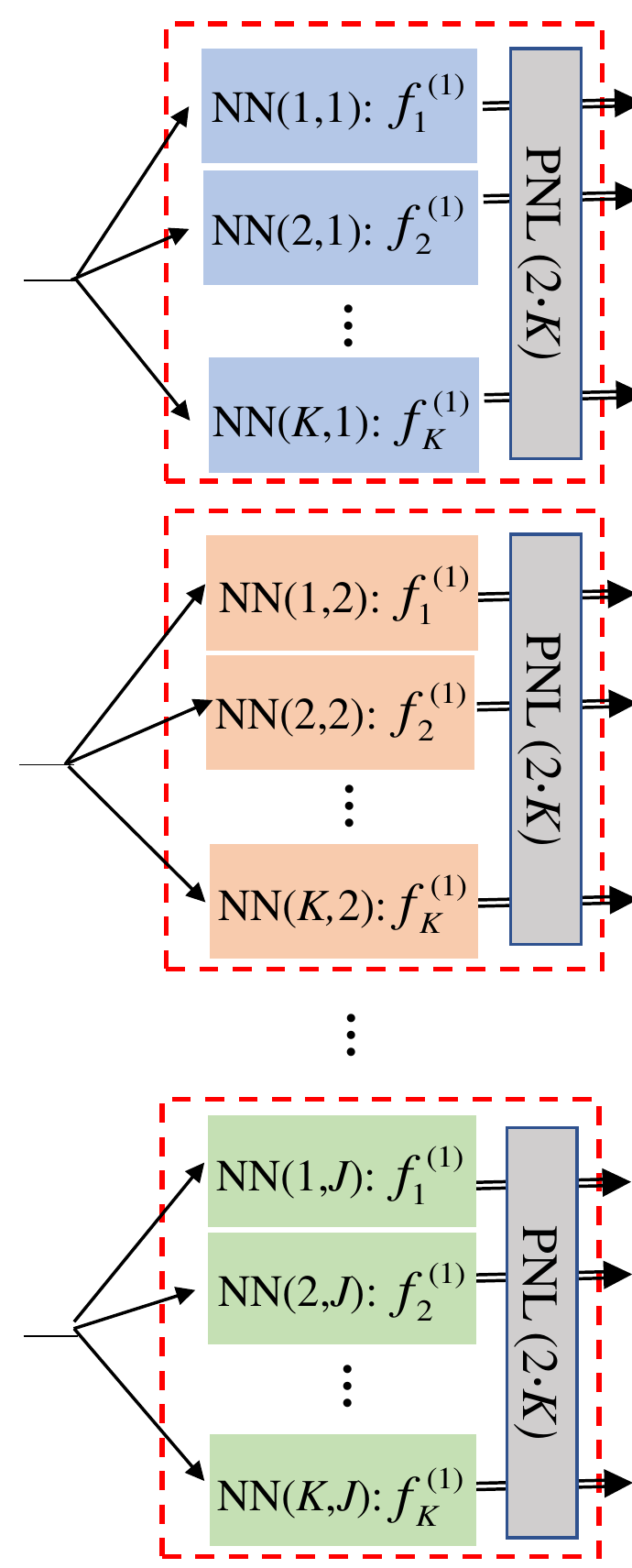}
\label{fig_sim}
\captionsetup{font=small,justification=centering}
  \caption{Level-2 PNL} 
\end{subfigure}
\begin{subfigure}[b]{0.38\linewidth}
\centering
\includegraphics[height=2.5in]{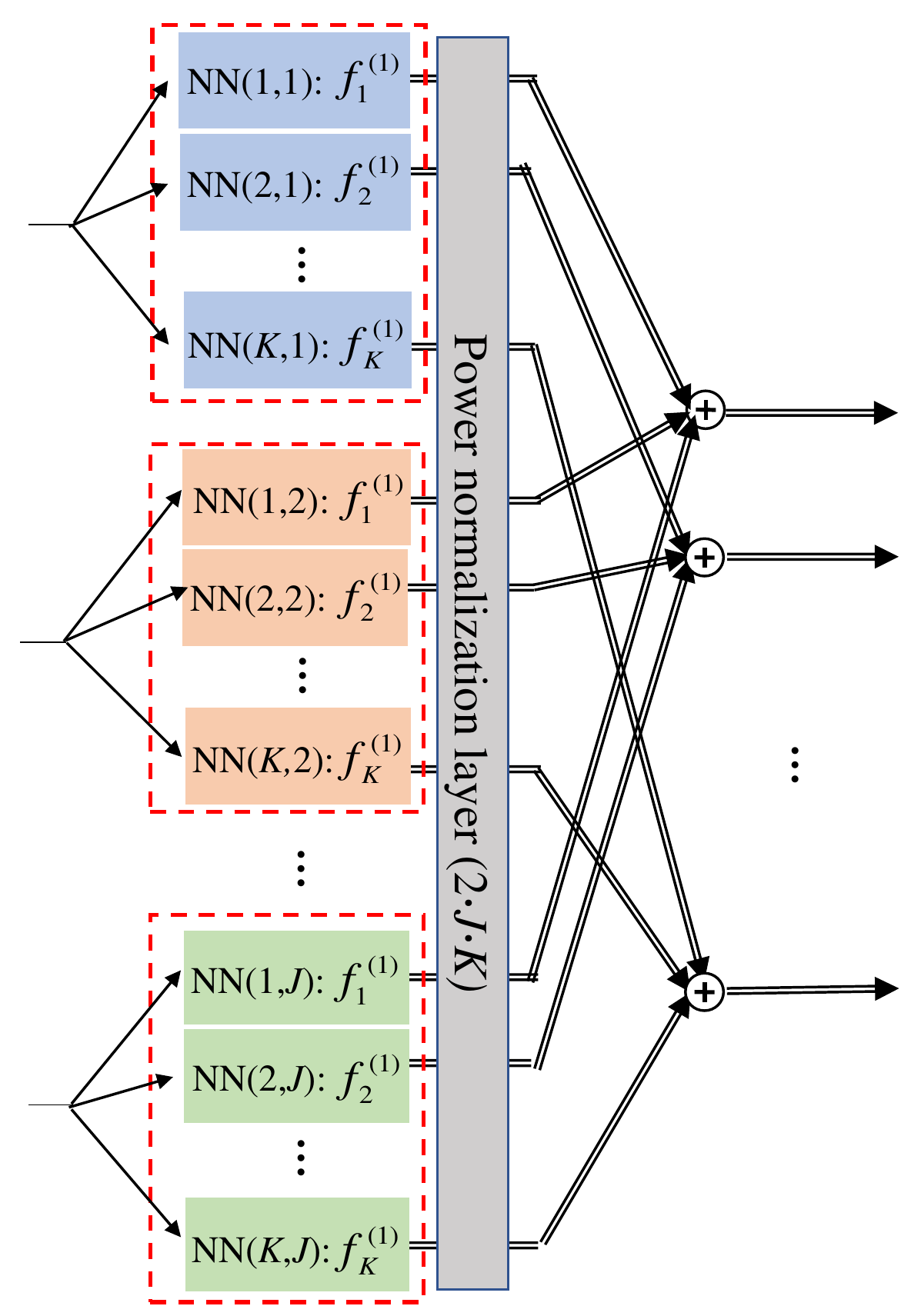}
\label{fig_sim}
\captionsetup{font=small,justification=centering}
  \caption{Level-3 PNL} 
\end{subfigure}
  \captionsetup{font=small,justification=raggedright, singlelinecheck=false}
\caption{Power normalization layers for the proposed MU-MDM encoder}
\end{figure*}

\section{Channel Autoencoder for MU-MDM Design}
\subsection{Autoencoder Structure for MU-MDM Model}
In this subsection, an AE structure for an MU-MDM design is generalized using an extension of that for an SU-MDM from Section \uppercase\expandafter{\romannumeral3}; the same order of training complexity is ensured subject to the same bandwidth efficiency. Fig. 4(a) shows the overall structure of the proposed AE for the MU-MDM design, including the CW element generator in Fig. 4(b). There are two inherent (indispensable) features in the AE structure for MU-MDM, which differentiate it from that for SU-MDM. The first is the input vector split, where the one-hot vector input $\mathbf{r}\triangleq {{[{{({{\mathbf{r}}^{(1)}})}^{T}},{{({{\mathbf{r}}^{(2)}})}^{T}},\cdots ,{{({{\mathbf{r}}^{(J)}})}^{T}}]}^{\text{T}}}$ associated with the input-bit sequence $\mathbf{b}\triangleq {{[{{({{\mathbf{b}}^{(1)}})}^{T}},{{({{\mathbf{b}}^{(2)}})}^{T}},\cdots ,{{({{\mathbf{b}}^{(J)}})}^{T}}]}^{T}}$ is compartmentalized into independent inputs, that is, $\{{{\mathbf{r}}^{(j)}}\}_{j=1}^{J}$. Using an input vector split, user $j$’s constellation ${{\mathsf{\pazocal{C}}}^{(j)}}=\{\mathbf{c}_{1}^{(j)},\mathbf{c}_{2}^{(j)},\cdots ,\mathbf{c}_{M}^{(j)}\}$ is determined by the input vector ${{\mathbf{r}}^{(j)}}$ only and is unaffected by the other user’s input vectors $\{{{\mathbf{r}}^{(j')}}\}_{j'=1}^{J},j'\ne j$. As a result, user $j$’s constellation set has a size of $M$, as discussed for the CD-NOMA system model in Section \uppercase\expandafter{\romannumeral2}-A. The second feature is an independent CW generator ${{f}^{(j)}}$ for each user, which generates the multi-dimensional CW over $K$ resources, as shown in Fig. 4(a). It should be noted that the CW generator ${{f}^{(j)}}$ is a concatenation of $K$ CW element generators, $\{f_{k}^{(j)}\}_{k=1}^{K}$, as given in Fig. 4(b).

Let ${{D}^{({{f}_{MU}})}}$ and ${{L}^{({{f}_{MU}})}}$ represent the hyperparameters used to denote the numbers of nodes and layers in $f_{k}^{(j)}$ on the encoder side, respectively. Similarly, ${{D}^{({{g}_{MU}})}}$ and ${{L}^{({{g}_{MU}})}}$ are hyperparameters that denote the numbers of nodes and layers on the decoder side, respectively. In the CW element generator $f_{k}^{(j)}$, there are two outputs for each dimension: one for the real part and another for the imaginary part of the complex signal points. The remaining the structure for the NN encoder can be determined by resource mapping vector $\mathbf{F}$ and the PNL. If the $(j,k)$-th element of $\mathbf{F}$ is 0, the PNL in $f_{k}^{(j)}$ forces the normalized power to be zero; otherwise, power is normalized with a given power constraint as prescribed in Section \uppercase\expandafter{\romannumeral2}-B.

Whenever one of the $M$ realizations for the one-hot sub-vector $\mathbf{r}_{i}^{(j)}$ is applied to the CW element generator $f_{k}^{(j)}$, it produces a complex signal point ${{\mathbf{s}}^{(j)}}$ in the $k$-th dimension. The $K$ complex dimensional signal points from $\{f_{k}^{(j)}\}_{k=1}^{K}$ that belong to each user are then superposed to form a constellation over $K$ complex dimensions, that is, $\mathbf{x}=\sum\nolimits_{j=1}^{J}{{{\mathbf{s}}^{(j)}}}$. For the corruption layer, a complex AWGN corruption vector $\mathbf{n}\sim \pazocal{C}\pazocal{N}(0,\sigma {{\mathbf{I}}_{K}})$ is then added to yield an intermediate signal $\mathbf{y}$. The corruption level must be carefully chosen because it determines the training and test performance of the AE [17]. The superposed and corrupted multi-dimensional constellation is fed into the NN-MUD, ${{g}_{MU}}(\cdot )$. The NN-MUD then produces an output vector $\mathbf{\hat{r}}={{[{{({{\mathbf{\hat{r}}}^{(1)}})}^{T}},{{({{\mathbf{\hat{r}}}^{(2)}})}^{T}},\cdots ,{{({{\mathbf{\hat{r}}}^{(J)}})}^{T}}]}^{\text{T}}}\in {{\mathbb{B}}^{1\times (J\cdot M)}}$ , which can be converted into the corresponding bit sequence, $\mathbf{\hat{b}}={{[{{({{\mathbf{\hat{b}}}^{(1)}})}^{T}},{{({{\mathbf{\hat{b}}}^{(2)}})}^{T}},\cdots ,{{({{\mathbf{\hat{b}}}^{(J)}})}^{T}}]}^{\text{T}}}\in {{\mathbb{B}}^{1\times J{{\log }_{2}}M}}$. Given the AE structure, an appropriate loss function must be considered to solve the optimization problem in (3).

Notably the input vector $\mathbf{r}$ with dimensions of $MJ$ result in ${{M}^{J}}$ different superposed signal points $\mathbf{x}$ in MU-MDM. Moreover, the AE for the SU-MDM in Fig. 3 with ${{M}^{J}}$ different signal points features the same input dimensions as that for the MU-MDM. This confirms that both the SU-MDM and MU-MDM schemes have the same number of bits per resource, that is, $(J{{\log }_{2}}M)/K$, while maintaining the same input dimensionality. In addition, all the layers and nodes in the AEs for the SU-MDM and MU-MDM are adjusted for a comparable number of training parameters; this ensures that the SU-MDM and MU-MDM have the same order of training complexity.

\subsection{Power Normalization Layer for MU-MDM Design}

Before discussing the different levels of PNL that can be applied to the AE, we note that an input vector is divided into separate sub-vectors that are independent of each other in the MU-MDM, as shown in Fig. 4(a). In this structure, all the generated CWs are constrained to the same average CB power in previous studies, i.e., [17] and [18]; therefore, this becomes a stringent design constraint as compared to SU-MDM. To reduce the performance gap between the MU-MDM and the baseline SU-MDM, we introduce a relaxed power normalization among the power assigned to the different CBs in this study. It is also worth mentioning that the relaxation in the power allocation to the different CBs has some similarities with the design of the PDMA [14] and PD-NOMA [2]. To evaluate the gain from the power constraint relaxation, we discuss three different levels of PNL, referred to as Level-1, Level-2, and Level-3 PNL, depending on the degree of freedom for the power allocation. 

Level-1 PNL addresses a CW element-wise constraint, which requires each non-zero element of a CW to be transmitted with the same power, as shown in Fig. 5(a). In this design, the CW element power normalization is given as follows:
\begin{align}
\left\| {{\mathbf{s}}^{(j)}}(k) \right\|_{2}^{2}=\left\{ \begin{matrix}
   \frac{{{P}_{j}}}{N}\text{  if  }f_{k}^{(j)}=1  \\
   0\text{  otherwise}  \\
\end{matrix} \right.,\forall j,k,
\end{align} 
where ${{\mathbf{s}}^{(j)}}(k)$ represents the $k$-th row of entry of CW vector ${{\mathbf{s}}^{(j)}}$. In Level-2 PNL, we consider CW-wise power normalization, that is, each CW is set to a fixed power, as shown in Fig. 5(b). The CW-wise power normalization is then given as
\begin{align}
\left\| {{s}^{(j)}} \right\|_{2}^{2}={{P}^{(j)}},\forall j.
\end{align} 
It should be noted that this is less stringent than element-wise (Level-1) power normalization, because each element (dimension) can be set to different powers. Therefore, it has a more room for optimization of constellation shape than element-wise optimization.

Lastly, in Level-3 PNL, the sum of the CW power is normalized, as shown in Fig. 5(c). This allows the CBs assigned to different users to have different powers, and they are only subject to a total sum constraint as follows: 
\begin{align}
\sum\limits_{j=1}^{J}{\left\| {{\mathbf{s}}^{(j)}} \right\|_{2}^{2}}=\sum\limits_{j=1}^{J}{{{P}^{(j)}}}=JP.
\end{align} 
The Level-3 design involves the least stringent constraint, as opposed to (9) and (10), which have the same average power for each user’s CB. This provides additional degrees of freedom for the CB design, achieving a higher shaping gain. In other words, the less stringent power constraint in Level-3 normalization enables the AE to increase the shaping gain, which, in turn, maximizes the minimum ED among multi-dimensional constellations. 

We note that the superior performance with a higher level of PNL has also been demonstrated in conventional CB designs for SCMA [10], [13]. In Table \uppercase\expandafter{\romannumeral1}, the different levels of PNLs are compared for the proposed and conventional design approaches. In the very first CB design for SCMA in [7], signal points that overlap in the same resource are differentiated by employing a rotated QPSK constellation in each dimension of CW. Consequently, each of the overlapped signal points in the CW element has the same power with a different phase as in our Level-1 PNL. The CB in [10] exploits a layer-specific operation, such as constellation rotation, with regard to the lattice-based mother constellation. Thus, it allows for power imbalances across the dimensions of CWs, resulting power differentiation among the colliding layers. However, its shaping gain is still limited because its mother constellation forces each user to have a fixed average CB power as in our Level-2 PNL. Recently in [13], a more flexible form of power-imbalanced CB design was proposed to “heuristically” optimize the average power allocated to different CBs. The BER performance can be significantly improved, as in our Level-3 PNL, owing to its higher multi-dimensional shaping gain, which is associated with a higher degree of freedom by relaxing a power constraint. 

However, besides the principle of power normalization, the proposed design framework is significantly different from that in [13], because we exploit the AE’s ability to optimize the multi-user multi-dimensional constellation in a given relaxed PNL, while considering bit-to-symbol mapping for BER minimization, which is be discussed in the following subsection. 

\begin{center}
\begin{table}[htbp]
\captionsetup{font=small,justification=centering}
\caption{Different Levels of CB Power Normalization: Comparison}

\begin{tabular}{| >{\centering\arraybackslash}m{2.1cm} | >{\centering\arraybackslash\columncolor[gray]{0.8}}m{1.85cm}| >{\centering\arraybackslash}m{1.22cm}| >{\centering\arraybackslash}m{1.9cm} |} \hline %

\textbf{Conventional Designs}&\textbf{DL-based Design}&\textbf{Average CB Power}&\textbf{Power Allocation to CW Eelements} \\ \hline
Vanilla SCMA [7] & Level-1: CW Element Power Normalization & Same & Same \\ \hline
Power-balanced CB Design for SCMA [10] & Level-2: CW-wise Power Normalization & Same & Different \\ \hline
Power-imbalanced CB Design for SCMA [13] & Level-3: Sum CW Power Normalization & Different & Different \\ \hline
\end{tabular}

\label{tab1}
\end{table}
\end{center}

\subsection{Proposed Loss Function and Training Procedure for Joint Optimization}
To explain the proposed loss function, we observe the factors governing the BER in the arbitrary multi-dimensional constellation [28]. Let ${{d}_{H}}({{\mathbf{b}}^{(i)}},{{\mathbf{b}}^{(j)}})$ denote the HD between ${{\mathbf{b}}^{(i)}}$ and ${{\mathbf{b}}^{(j)}}$. It should be noted that a pairwise symbol error probability (PEP) ${{p}_{s}}(\cdot )$ is determined by the ED of the $i$-th and $j$-th CWs, which are given as $f({{\mathbf{r}}^{(i)}})$ and $f({{\mathbf{r}}^{(j)}})$. Let ${{\varepsilon }_{b}}$ denote the average BER for arbitrary ${{M}^{J}}$ multi-dimensional signal points, which can be computed as
\begin{align}
{{\varepsilon }_{b}}=\frac{1}{J{{\log }_{2}}M}\sum\limits_{i=1}^{{{M}^{J}}}{\sum\limits_{j\ne i}{{{d}_{H}}({{\mathbf{b}}^{(i)}},{{\mathbf{b}}^{(j)}})\cdot {{p}_{s}}\left( {{\mathbf{r}}^{(i)}}\to {{\mathbf{r}}^{(j)}} \right)}}.
\end{align} 
We first observe from (12) that the BER is governed by the multiplicative relationship between the PEP of the CWs and its corresponding HD. The PEP is a function of the SNR; therefore, the optimal multiplicative relationship between ED and HD also depends on the SNR level. However, there is no explicit expression that captures the multiplicative relationship in ED and HD to minimize the BER in terms of the SNR level. Therefore, the BER in (12) cannot be directly employed as an objective function for constellation design to minimize the BER.

We note that the conventional design involves two separate steps: the first step comprises a constellation design to minimize the CW error and the second step comprises bit-to-symbol mapping to minimize the BER for the given constellation, such as via Gray mapping. However, in the proposed MU-MDM AE design, a loss function must be designed to jointly optimize the signal points for the MU-MDM constellation and their bit-to-symbol mapping. Toward that end, when a symbol has large HD from the adjacent symbols, a large L2-norm should be applied to that symbol to increase the ED from those adjacent ones. In other words, a larger L2-norm loss should be applied to symbols with a larger Hamming error ${{d}_{H}}({{\mathbf{b}}^{(j)}},{{\mathbf{\hat{b}}}^{(j)}})$, which corresponds to the HD between the input and output bit sequences. Therefore, the loss function can be obtained by averaging the L2-norm for all possible CWs, each weighted by the Hamming error. 

However, as discussed previously, the appropriate amount of Hamming error weighted with the L2-norm in minimizing BER depends on the SNR. To adjust the weight based on the Hamming error such that a minimal BER can be maintained as far as possible even under a varying SNR, a weight constant, denoted by $\delta$, is introduced as a new hyperparameter to ensure that the design function over a wide range of SNR levels. Subsequently, for a given Hamming error, the following weight function with the hyperparameter $\delta $ is defined for user $j$:
\begin{align}
B({{\mathbf{b}}^{(j)}},{{\mathbf{\hat{b}}}^{(j)}})=\mu +\delta {{d}_{H}}({{\mathbf{b}}^{(j)}},{{\mathbf{\hat{b}}}^{(j)}}),
\end{align}
where $\mu $ is a non-zero positive constant that prevents a loss function from becoming zero in the correct detection case [16].

Thereafter, we attempt to minimize the average CW error, which is weighted by (14). In other words, the following proposed loss function, denoted by ${{L}_{proposed}}$, can be defined: 
\begin{align}
{{L}_{proposed}}=\frac{1}{J}\sum\limits_{j=1}^{J}{B({{\mathbf{b}}^{(j)}},{{{\mathbf{\hat{b}}}}^{(j)}})}\cdot {{\left\| {{\mathbf{r}}^{(j)}}-{{{\mathbf{\hat{r}}}}^{(j)}} \right\|}_{2}}.
\end{align}
Note that ${{\theta }^{(e)}}$ and ${{\theta }^{(d)}}$ are updated to minimize the loss function (14), in which the ED between CWs is mutually balanced with the HD between the input and output sequences; thus, their combined effect can be minimized. When $\delta $ approaches zero in (14), the loss function in (14) converges to the ED-based L2-norm loss function, which is favorable for a high SNR, and the Hamming error is neglected. However, a large value of $\delta $ deviates from the maximum ED constellation, which is favorable for a low SNR, and the HD between adjacent signal points in the constellation is minimized. This implies that there exists an appropriate value of the hyperparameter $\delta $ for the given SNR level. 

However, according to Proposition 1, deviations from the maximum ED constellation caused by $\delta $ may induce further CW errors, i.e., it may result in an unfavorable BER performance in the high SNR region. Thus, we switch the loss function in (14) to the L2-norm after sufficient convergence because the impact of Hamming weight in the loss function (14) is significant during the initial stage of training [16]. After the bit-to-symbol mapping is completed using the loss function (14), the effect of a constellation deviation from $\delta $ can be reduced with the maximum ED constellation. Therefore, the AE can be further trained using the L2-norm loss function after sufficient convergence with the loss function (14). Two different loss functions can be employed for the low and high SNR ranges; therefore, this constitutes a two-step training process, which allows AE to achieve the maximum ED constellation with near-optimal bit-to-symbol mapping. Once a signal constellation is optimized to reduce the BER in Step 1, its performance can be further improved by maximizing the ED between the bit-to-symbol-mapped signal points in Step 2. 

\begin{figure}[t]
\centering
\includegraphics[width=2.7in,height=2.3in]{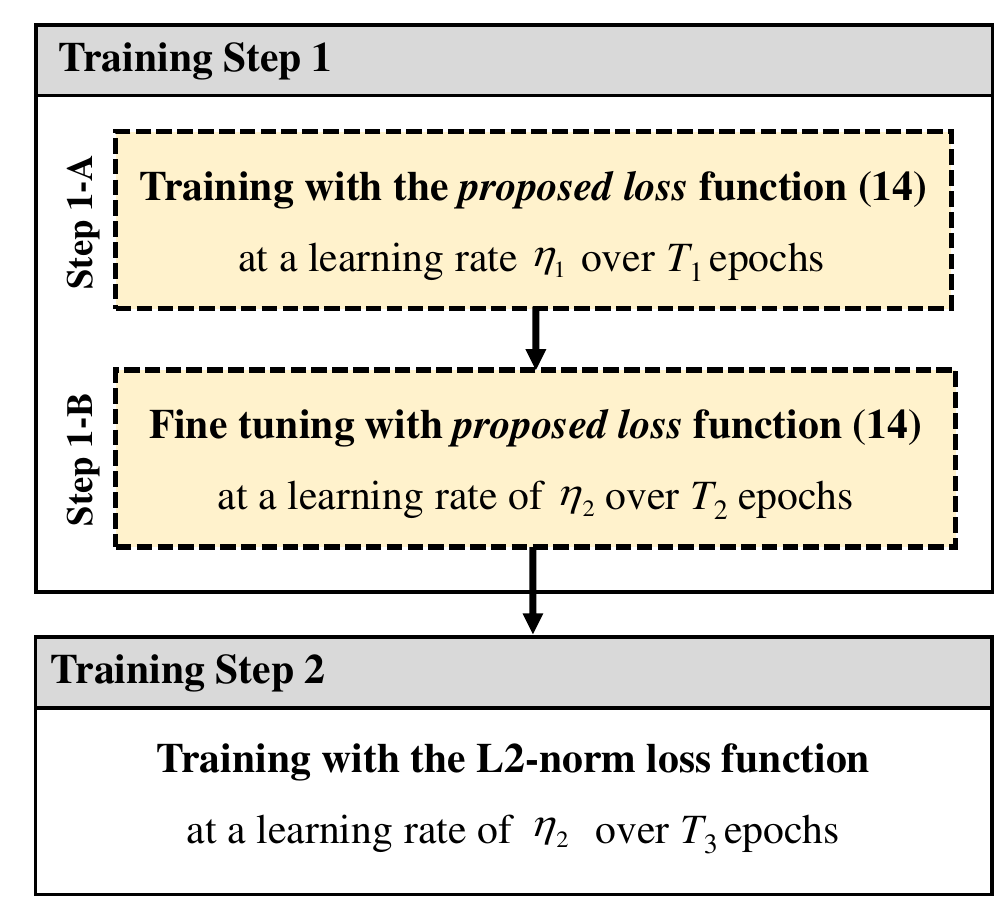}
  \captionsetup{font=small,justification=raggedright, singlelinecheck=false}
\caption{Proposed two-step training process}
\label{fig_sim}
\end{figure}

Two-step training process is depicted in Fig. 6. In Training Step 1, the proposed NN-based AE is trained with a learning rate of ${{\eta }_{1}}$ until both the training loss and the validation loss sufficiently converges within ${{T}_{1}}$ epochs (Step 1-A). After the loss hits a plateau, the learning rate is reduced to ${{\eta }_{2}}$, i.e., ${{\eta }_{2}}<{{\eta }_{1}}$, which allows for an increase in the convergence rate. After the network is fine-tuned with ${{\eta }_{2}}$ over ${{T}_{2}}$ epochs (Step 1-B), Training Step 2 is initiated by switching the loss function to L2-norm. Subsequently, it is trained over ${{T}_{3}}$ epochs until it converges. All the epochs, ${{T}_{1}}$, ${{T}_{2}}$, and ${{T}_{3}}$, can be set to be sufficiently large in order to ensure convergence. However, the underlying convergence rate is not critical in practice, simply because the proposed AE is trained offline; only the trained encoder–decoder pair is exploited online.

\section{Simulation Results}
In this section, the performance improvement achieved by proposed scheme, as compared to existing conventional and DL-based constellation designs, is demonstrated through extensive simulation results. To explain the BER performance achieved by the proposed scheme, we also present the process of hyperparameter optimization, while interpreting the constellation diagram obtained by the current design.

\subsection{Simulation Setup}

 In our simulation, we consider a CD-NOMA system with $J=6$ users, $K=4$ resources, and a modulation order of $M=4$, as described in many existing studies. We also consider both sparse ($N<K$) and dense ($N=K$) resource-mapping matrices. First, for the sparse mapping case, we consider the following resource mapping matrix:
\begin{align}
\mathbf{F}=\left[ \begin{array}{*{35}{l}}
   0 & 1 & 1 & 0 & 0 & 1  \\
   1 & 0 & 1 & 0 & 1 & 0  \\
   1 & 0 & 0 & 1 & 0 & 1  \\
   0 & 1 & 0 & 1 & 1 & 0  \\
\end{array} \right].
\end{align}                     
For the dense mapping, we consider dense resource mapping matrix where all the entries are one. Detailed AE structures corresponding to the sparse and dense resource mapping matrices are described in Section \uppercase\expandafter{\romannumeral4}-A. 

We consider the offline training of the AE with the proposed structure and the loss function. A set of output CBs, ${{\mathsf{\pazocal{S}}}^{(j)}}=\{\mathbf{s}_{1}^{(j)},\mathbf{s}_{2}^{(j)},\cdots ,\mathbf{s}_{J}^{(j)}\}$, is generated with $\{{{f}^{(j)}}\}_{j=1}^{J}$ for all possible input vectors. Furthermore, we directly employ the trained decoder function $g(\cdot )$ as an NN-MUD receiver, which achieves approximately the same BER performance as the MPA receiver while incurring a much lower complexity [17]-[19].

We generate a dataset with entire data classes for the dataset composition of the offline training. A dataset of ${{M}^{J}}$ classes was constructed as input vectors $\mathbf{r}\in {{\mathbb{B}}^{MJ}}$. Because random corruption is added to the hidden layer, the data fed as the input to $g(\cdot )$ always vary, even in the case where it is set to the same input vector. This prevents the NN decoder from being overfitted to the dataset with a limited size; thus, it is generalized to the unseen channel realizations. Offline training is applied to obtain only the CB set $\pazocal{S}$ from the encoder function; therefore, we are not concerned about the system complexity associated with ${{D}^{({{f}_{MU}})}}$ and ${{L}^{({{f}_{MU}})}}$. Therefore, ${{D}^{({{f}_{MU}})}}$ and ${{L}^{({{f}_{MU}})}}$ can be appropriately selected so that NN can be neither overfitted nor underfitted in a given training scenario. However, the receiver complexity of the NN-MUD is directly governed by ${{D}^{({{g}_{MU}})}}$ and ${{L}^{({{g}_{MU}})}}$ in the actual system. Therefore, they must be minimized as much as possible, while being neither overfitted nor underfitted under a given training scenario.

\subsection{Hyperpatameter optimization}
The rectified linear unit (ReLU) is preferred over the hyperbolic tangent function (tanh) as an activation function for faster convergence, owing to a very deep NN with millions of trainable parameters in the proposed AE. However, zero-centered encoded data work better with using an activation function with a zero-centered shape. In fact, the tanh function is more appropriate for the constellation design, as shown in [16], [18]; its BER performance is better than that obtained with the ReLU. This is also true for the decoder side because the channel output is also composed of zero-centered data. Therefore, in the proposed AE, tanh activation functions are employed for both the encoder and the decoder. Moreover, a stochastic gradient descent optimizer is used with the given learning rate. 

The corruption level and weight constant $\delta $ for the loss function are two important hyperparameters that need to be optimized to implement the proposed AE network. The appropriate corruption level and $\delta $ depend on various parameters, such as the number of users, number of resources, CB power level of the user, and PNL level; therefore, extensive BER simulations with the Level-3 PNL are performed to select the appropriate hyperparameters that can minimize the BER under a general AWGN channel with various SNR level. Meanwhile, we note that a common practice in the conventional CD-NOMA is to design a constellation under AWGN channel and then, any channel can be converted into the AWGN channel through channel estimation and multi-user equalization [6]-[13]. We can also consider constellation rescaling scheme in [30], which generalizes the proposed constellation to the Rayleigh channel. However, generalization of the proposed constellation to the Rayleigh channel will be beyond a scope of the current work. 

The corruption level for each user is defined as the ratio of the AWGN power to the average CB power, ${{\sigma }^{2}}/E[{{P}^{(j)}}]$. If this value is excessively high, the AE fails to converge without learning the signal points. By contrast, if this value is excessively low, the AE converges to a local optimal point without fully exploring the globally optimal constellations; thus, the AE is overfitted to the training dataset. This suggests that there must exist an appropriate level of corruption, as discussed in [17]. In this study, the appropriate corruption level in the MU-MDM design is identified as $-$8 and $-$6 dB in the sparse and dense $\mathbf{F}$, respectively, based on extensive BER simulations. In other words, the $-$8 dB and $-$6 dB corruption levels generalize the low SNR region with fair convergence under the sparse and dense settings, respectively. The appropriate corruption level in dense CD-NOMA is slightly higher than that in sparse CD-NOMA, which is attributed to the larger average inter-CW distance in the dense resource mapping. This enables the for dense CD-NOMA to be more resilient against AWGN.

\begin{figure}[t]
\centering
\includegraphics[width=3in,height=2.25in]{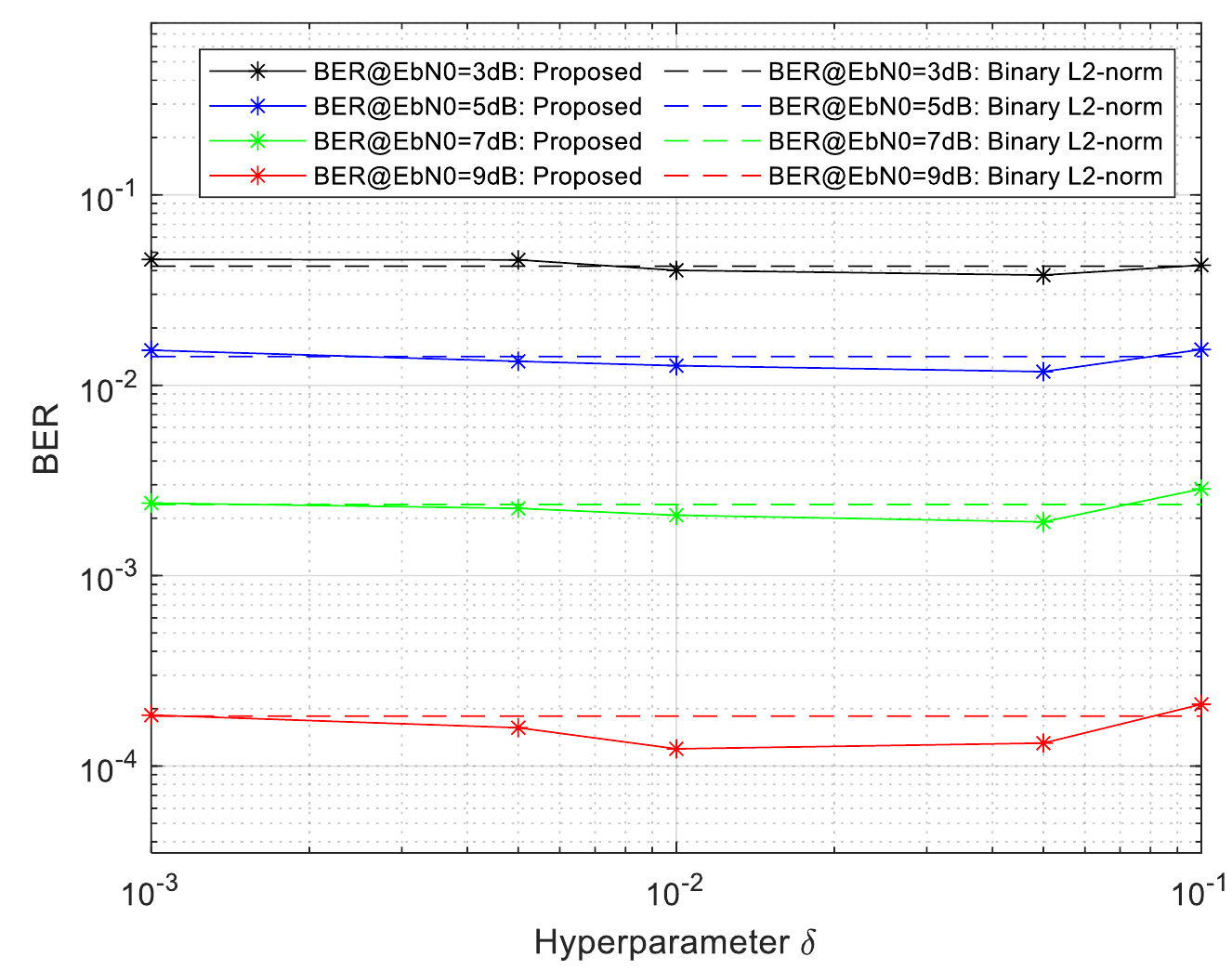}
  \captionsetup{font=small,justification=raggedright, singlelinecheck=false}
\caption{BER performance vs. weight constant $\delta$}
\label{fig_sim}
\end{figure}

In Fig. 7, we present the performance results that are used to select the useful weight constant $\delta$ in the proposed loss function. It shows the BER  performance with the proposed loss function varying with respect to the weight constant $\delta$ as a hyperparameter for the given SNR. Its performance is compared with that obtained using the L2-norm loss function with binary vector representation (represented by the dotted lines in the graphs). It is observed that the proposed loss function generally outperforms the L2-norm loss function for a particular weight constant range, i.e., $0.01\le \delta \le 0.05$, from the low to high SNR levels, thereby fulfilling our design objectives. Therefore, in the performance analysis for the proposed design with dense resource mapping and Level-3 PNL, the weight constant $\delta$ is selected from this range, such as $\delta =0.05$. Detailed hyperparameters for the DL used in the current design are summarized in Table \uppercase\expandafter{\romannumeral2}.

\begin{table}[htbp]
\captionsetup{font=small,justification=centering}
\caption{Hyperparameters for deep learning: summary}
\begin{center}
\begin{tabular}{|M{1.98in}|M{0.45in}|M{0.5in}|} \hline
\textbf{Hyperparameters}&\textbf{Notation}&\textbf{Value} \\ \hline
Batch size & ${N}_{batch}$ & 400 \\ \hline
\multirow{2}*{Learning Rate} & $\eta_{1}$ & 0.001 \\ \cline{2-3}
& $\eta_{2}$ & 0.0001 \\ \hline
\multirow{3}*{Number of epochs} & $T_{1}$ & 8000 \\ \cline{2-3}
& $T_{2}$ & 2000 \\ \cline{2-3}
& $T_{3}$ & 1000 \\ \hline
Corruption level & ${{\sigma }^{2}}/E[{{P}^{(j)}}]$ & $-$6 dB for dense and $-$8 dB for sparse \\ \hline
Weight constant in the proposed loss & $\delta$ & 0.05 \\ \hline 
Number of nodes in the MU-MDM CW element generator & ${{D}^{({{f}_{MU}})}}$ & 32 \\ \hline
Number of layers in the MU-MDM CW element generator & ${{L}^{({{f}_{MU}})}}$ & 6 \\ \hline
Number of nodes in the MU-MDM decoder & ${{D}^{({{g}_{MU}})}}$ & 512 \\ \hline
Number of layers in the MU-MDM decoder & ${{L}^{({{g}_{MU}})}}$ & 4 \\ \hline
Number of nodes in the SU-MDM encoder & ${{D}^{(f_{SU})}}$ & 256 \\ \hline
Number of layers in the SU-MDM encoder & ${{L}^{(f_{SU})}}$ & 6 \\ \hline
Number of nodes in the SU-MDM decoder & ${{D}^{(g_{SU})}}$ & 512 \\ \hline
Number of layers in the SU-MDM decoder & ${{L}^{(g_{SU})}}$ & 4 \\ \hline
\end{tabular}
\label{tab1}
\end{center}
\end{table}

\begin{figure*}[ht]
\centering
\includegraphics[width=0.75\linewidth,height=3.35in]{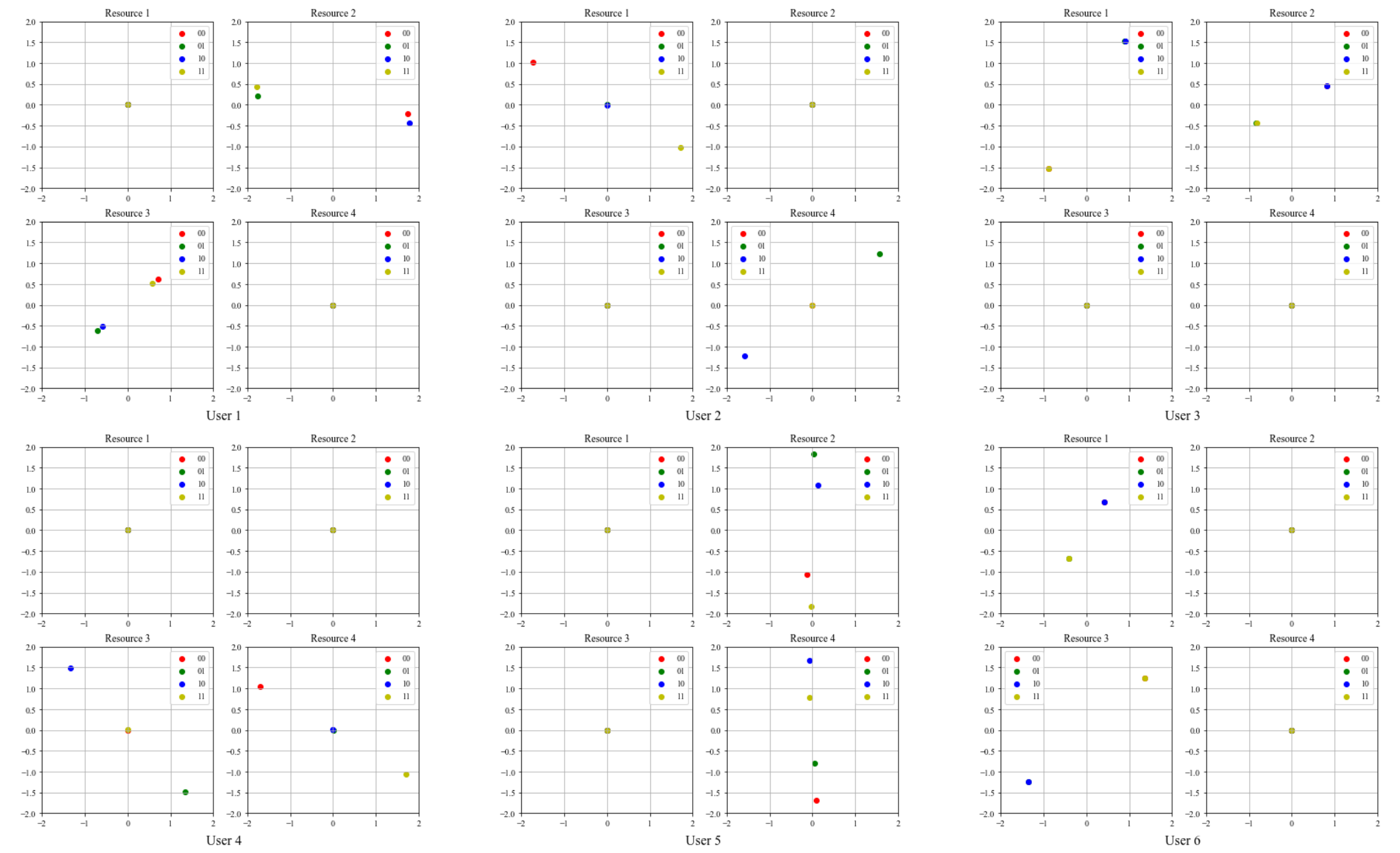}
  \captionsetup{font=small,justification=raggedright, singlelinecheck=false}
\caption{Constellation for sparse CD-NOMA with the sum power normalization constraints}
\label{fig_sim}
\end{figure*}

\begin{figure*}[t]
\centering
\begin{subfigure}[b]{0.32\linewidth}
\centering
\includegraphics[width=2.1in,height=1.85in]{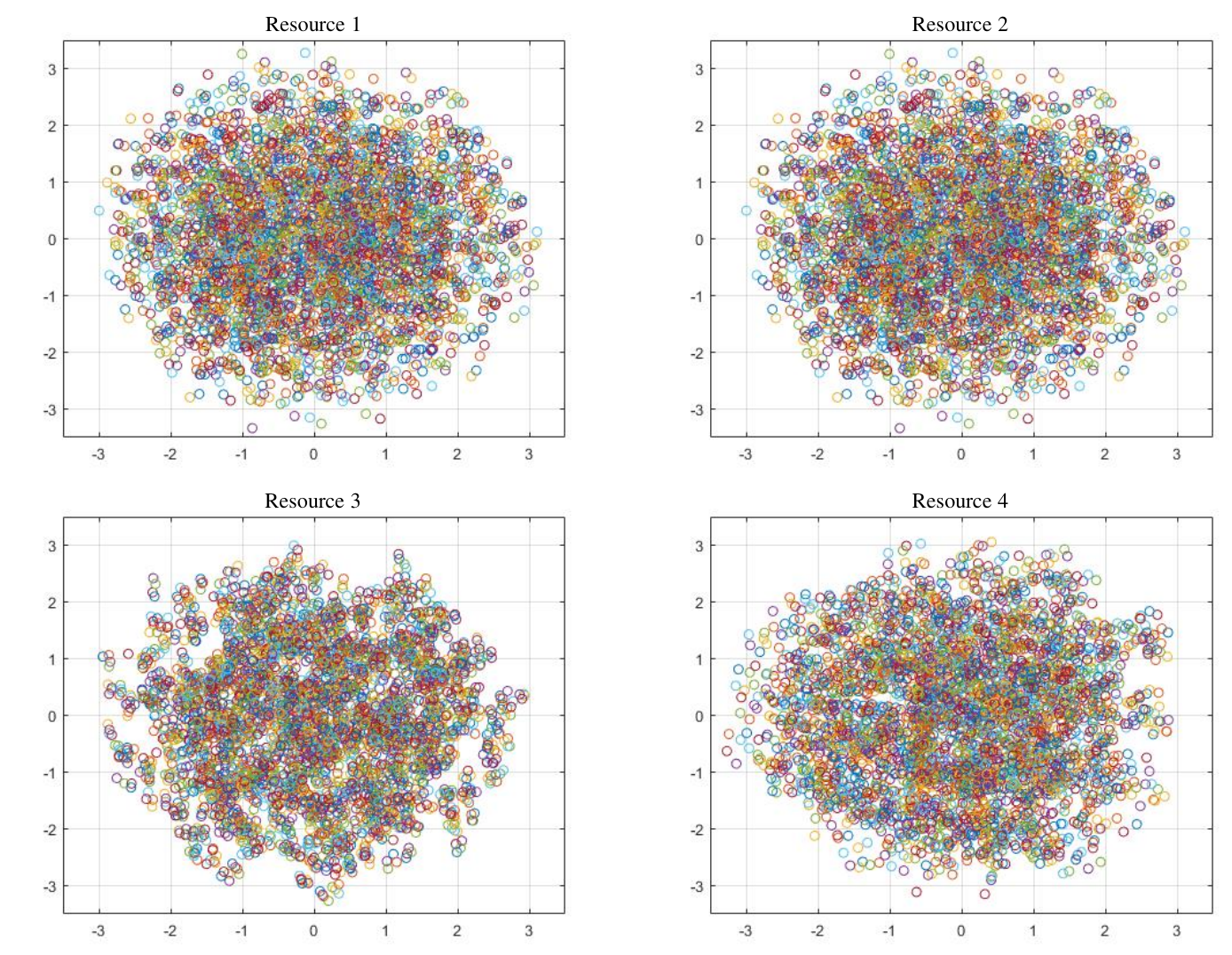}
\label{fig_sim}
\captionsetup{font=small,justification=centering}
  \caption{DL-based SU-MDM} 
\end{subfigure}
\begin{subfigure}[b]{0.32\linewidth}
\centering
\includegraphics[width=2.1in,height=1.85in]{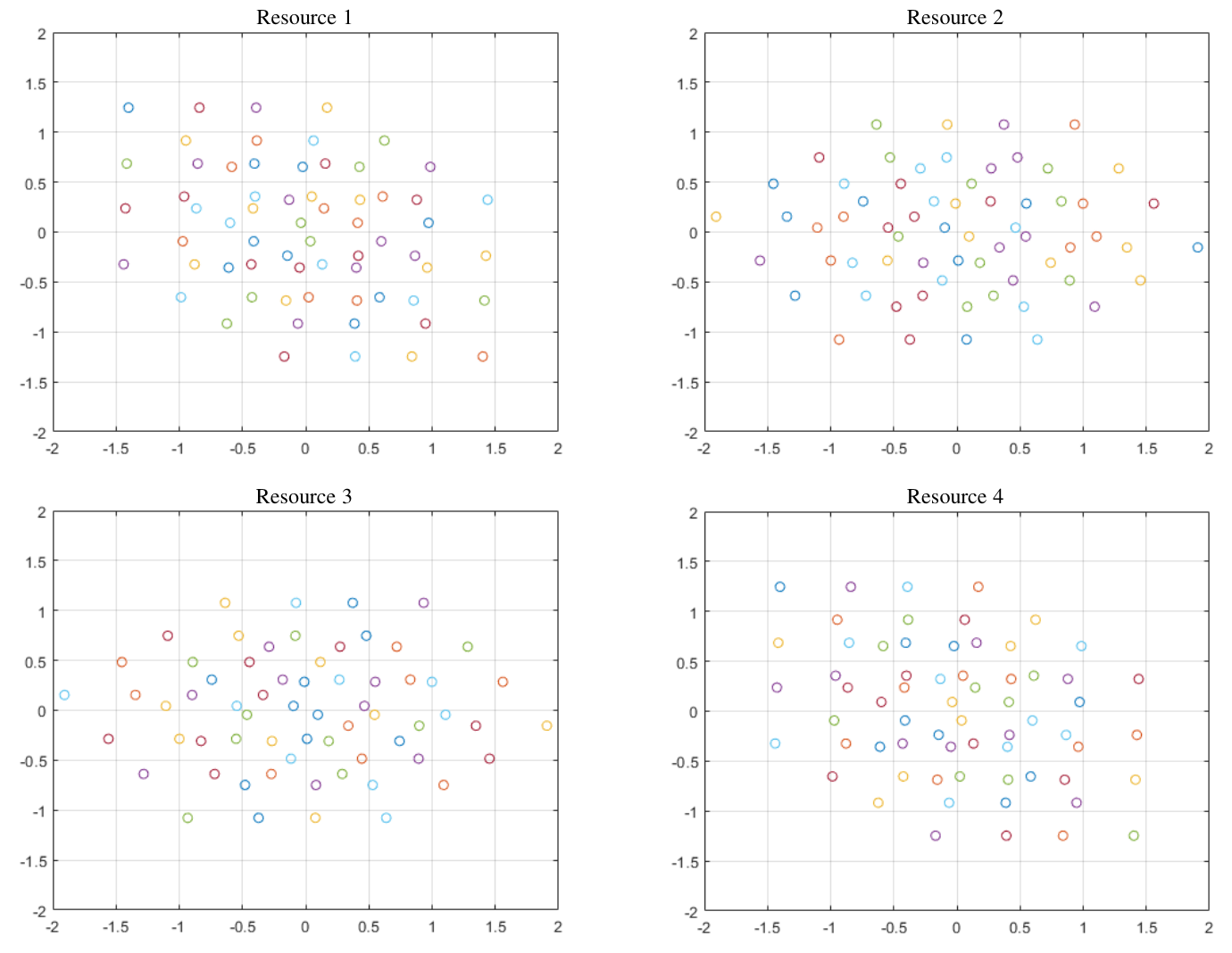}
\label{fig_sim}
\captionsetup{font=small,justification=centering}
  \caption{Conventional SCMA} 
\end{subfigure}
\begin{subfigure}[b]{0.32\linewidth}
\centering
\includegraphics[width=2.1in,height=1.85in]{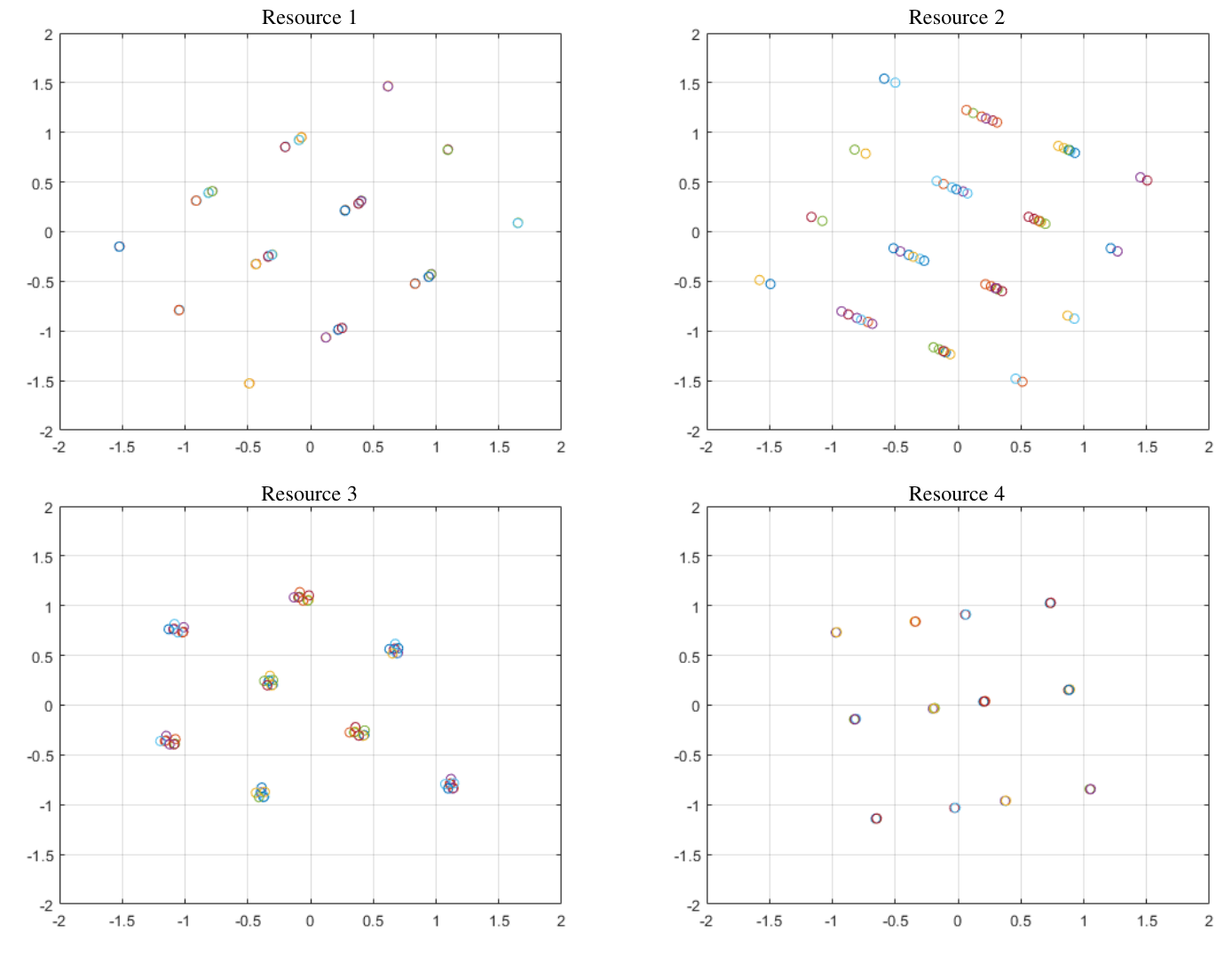}
\label{fig_sim}
\captionsetup{font=small,justification=centering}
  \caption{DL-based MU-MDM} 
\end{subfigure}
  \captionsetup{font=small,justification=raggedright, singlelinecheck=false}
\caption{Comparison of superposed constellations for the different designs}
\end{figure*}

\subsection{Constellation Diagrams}
Fig. 8 shows an example of the constellation diagrams for sparse CD-NOMA using F given in (15). These constellations are constructed using the proposed DL-based MU-MDM with Level-3 PNL. There are two distinctive features in the designed CBs, as compared with the conventional CBs. First, there are zero-power signal points in the allocated resources, such as for users 2 and 4. This implies that some users can use fewer resources than they are assigned, depending on which constellation that each user attempts to transmit. Even if the transmit diversity gain is reduced because some resources remain unused, the inter-user distance between the signal points can be increased to minimize the BER. Second, unlike the CB for the conventional SCMA, some constellations only have two signal points; this also results in underutilization of the resources. Notably, ${{\log }_{N}}M$ signal points per resource are sufficient to uniquely distinguish $M$ signal points when N resources are utilized by a user. Even if ${{\log }_{N}}M$ signal points per utilized resource are sufficient, it is significantly complex to design CBs in which a subset of the resources is utilized. However, DL allows for the learning of constellations by varying the number of constellations per resource, while maximizing the inter-constellation distance and minimizing the SER. Consequently, as shown in Fig. 8, three different types of constellations are observed for the different resources: $M$-symbol constellation, ${{\log }_{N}}M$-symbol constellation, and constellation with zero-powered symbols.

Fig. 9 shows the constellations for the various designs before the AWGN corruption is applied. Fig. 9(a) shows the DL-based SU-MDM constellation, which includes ${{M}^{J}}$ signal points in each resource; this represents the unique realization of the input vector. Although the constellation in Fig. 9(a) achieves good performance, it cannot be applied to MU-MDM constellations, as discussed in Subsection IV-A. Fig. 9(b) and 9(c) show the conventional SCMA-based superposed CB and the proposed DL-based superposed constellation, respectively. The signal points in the superposed constellation correspond to ${{M}^{J}}$ possible realizations of the vector $\mathbf{x}$, as described in Subsection II-A. It is straightforward for the CB of the conventional SCMA to feature unique ${{M}^{{{d}_{f}}}}$ superposed constellations for each resource. By contrast, the DL-based constellations exhibit a clustered form of the superposed constellations. We observe that the inter-cluster constellation distance is considerably larger than that for the conventional CBs; however, the ED within the superposed constellation cluster is extremely small.

\subsection{Performance Analysis}

\begin{figure}[t]
\centering
\begin{subfigure}[b]{\linewidth}
\centering
\includegraphics[width=3.1in,height=2.6in]{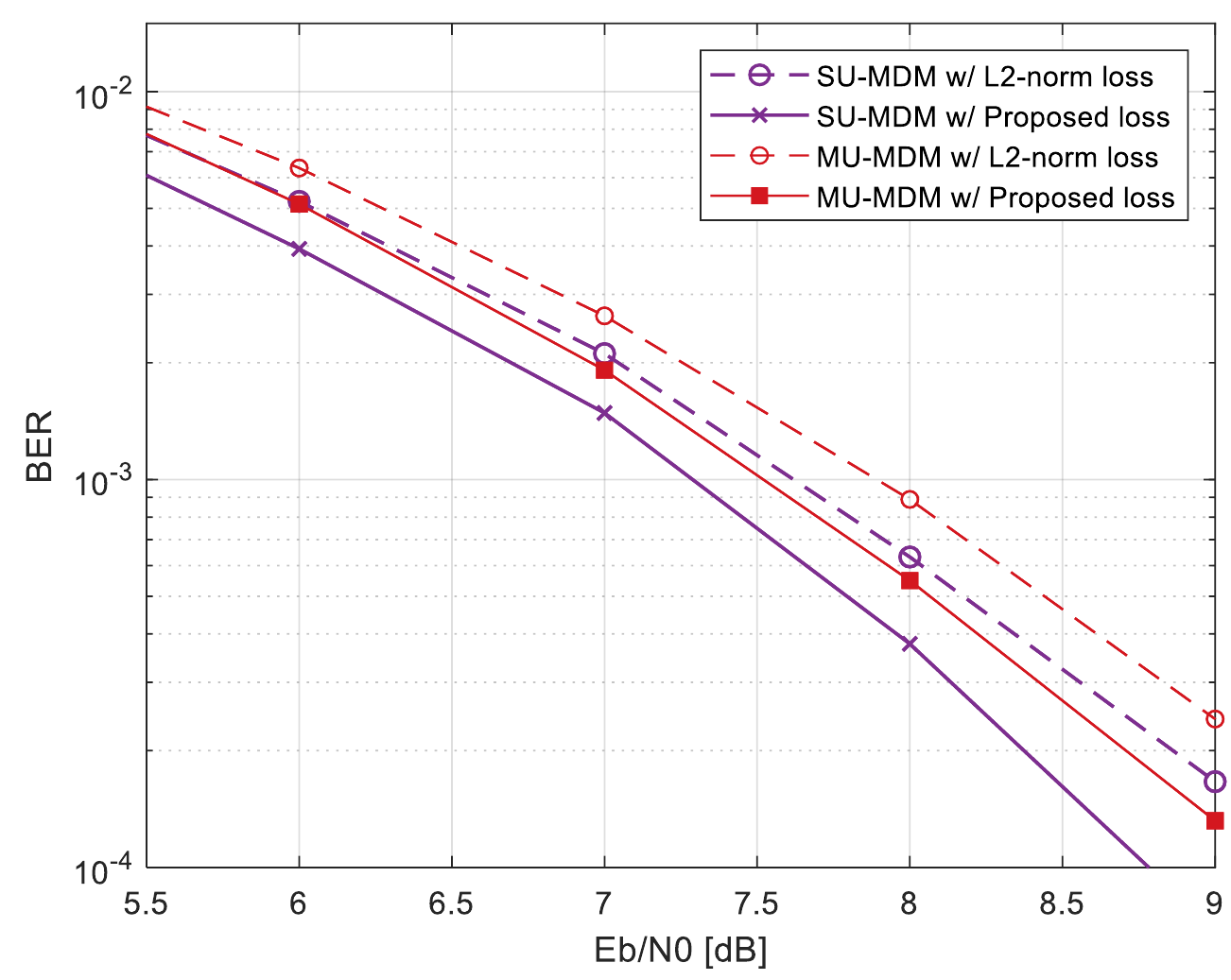}
\label{fig_sim}
\captionsetup{font=small,justification=centering}
  \caption{SU-MDM vs. MU-MDM}
\end{subfigure}
\begin{subfigure}[b]{\linewidth}
\centering
\includegraphics[width=3.1in,height=2.6in]{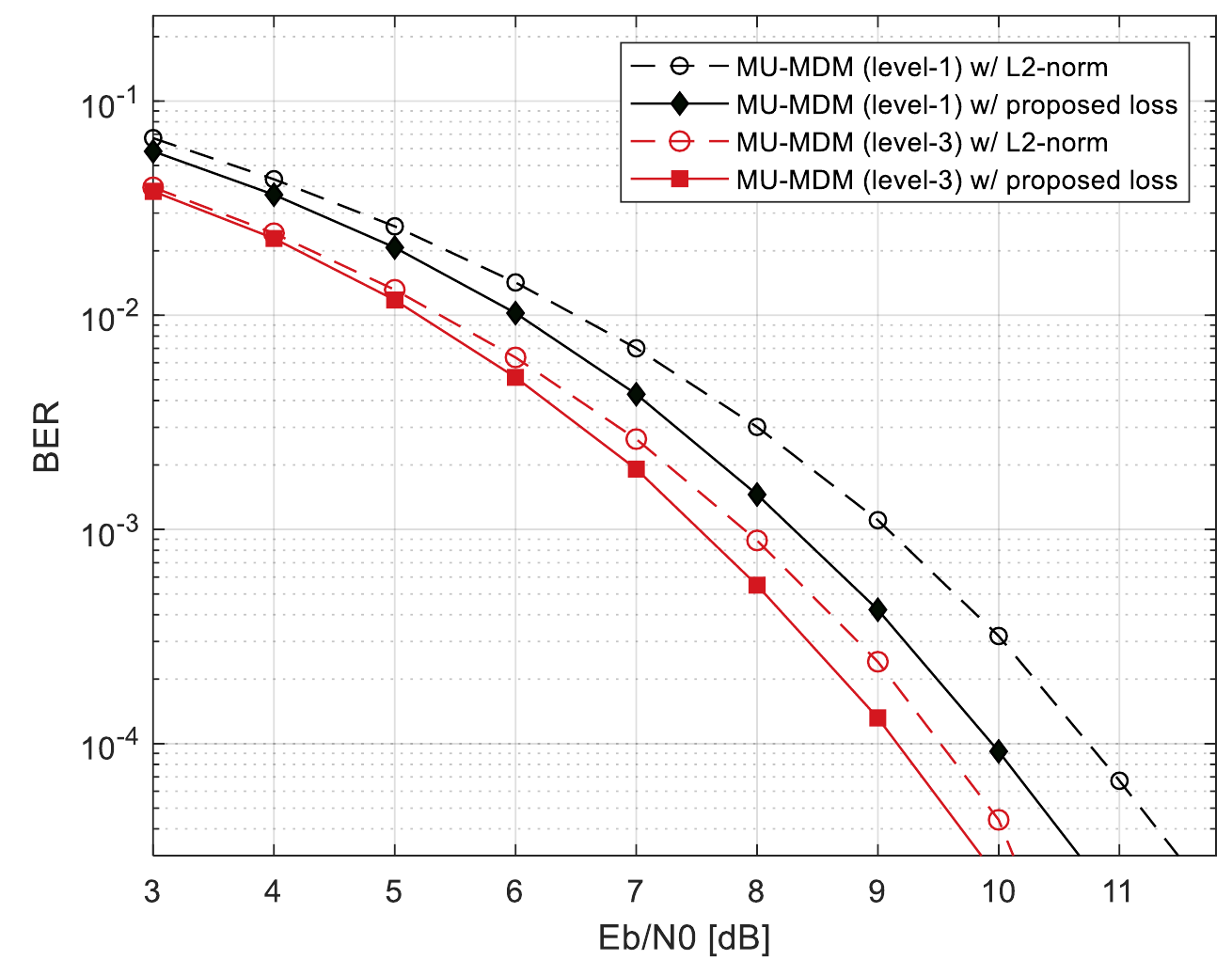}
\label{fig_sim}
\captionsetup{font=small,justification=centering}
  \caption{Level-1 vs. Level-3 PNL} 
\end{subfigure}
  \captionsetup{font=small,justification=raggedright, singlelinecheck=false}
\caption{Comparison of BER performance: L2-norm vs. the proposed loss function }
\end{figure}

\begin{figure}[t]
\centering
\includegraphics[width=3.1in,height=2.6in]{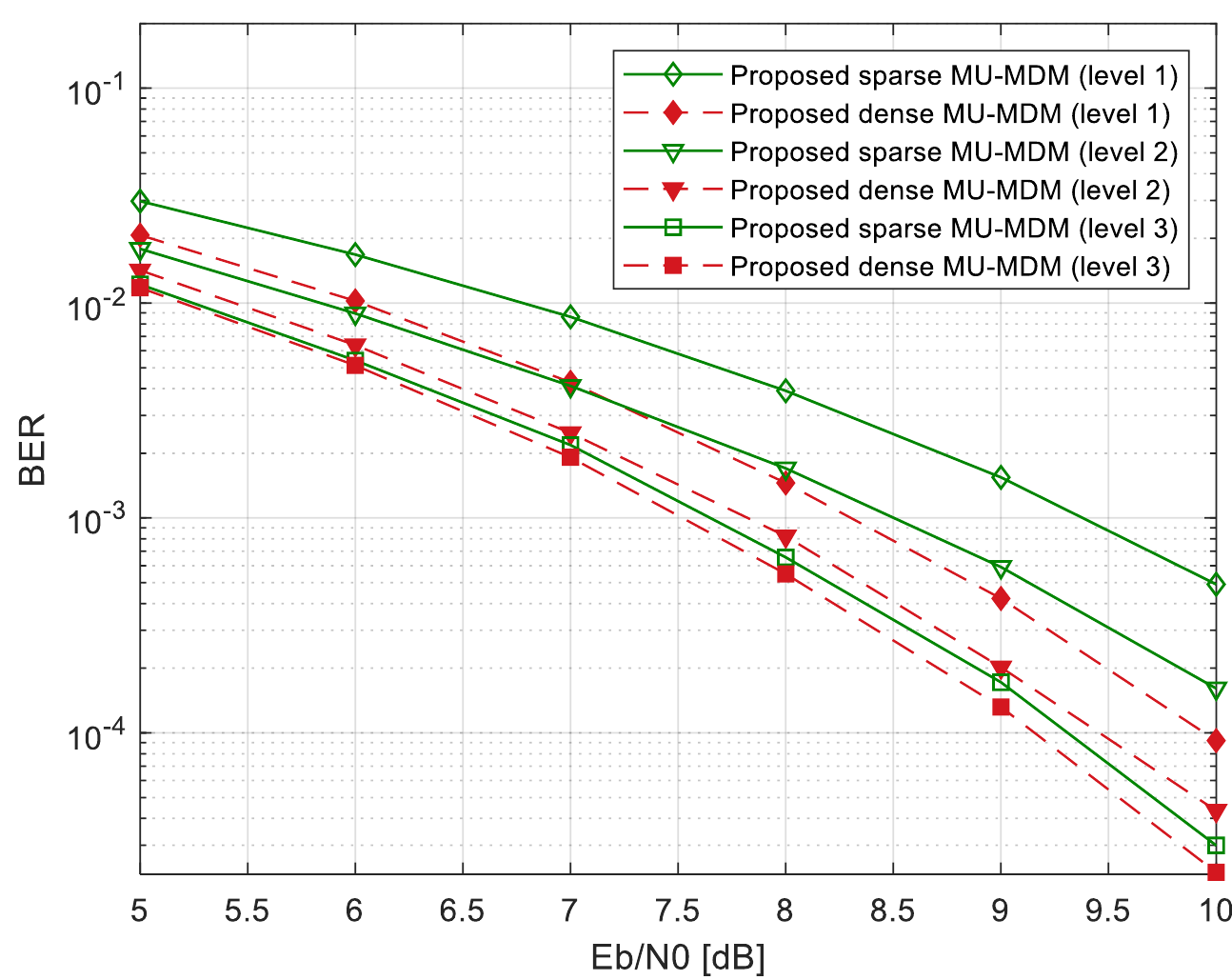}
  \captionsetup{font=small,justification=raggedright, singlelinecheck=false}
\caption{Comparison of BER performance with different levels of PNL: Dense vs. sparse resource mapping}
\label{fig_sim}
\end{figure}

\begin{figure}[t]
\centering
\includegraphics[width=3.1in,height=2.6in]{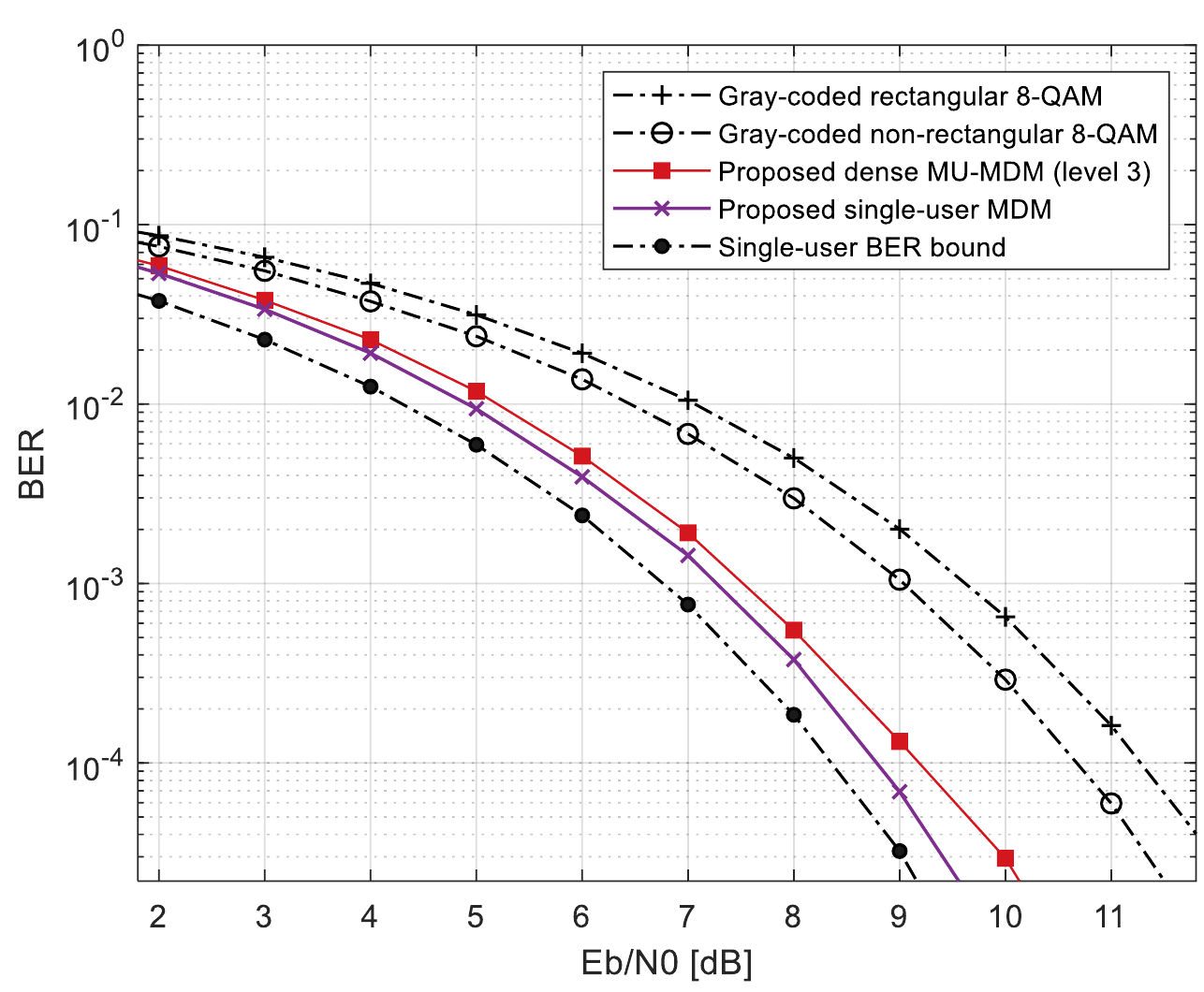}
  \captionsetup{font=small,justification=raggedright, singlelinecheck=false}
\caption{Comparison of BER performance: The proposed MU-MDM vs. proposed SU-MDM}
\label{fig_sim}
\end{figure}

First of all, we investigate how our proposed loss function outperforms the plain L2-norm loss function for the different setting in Fig. 10. Fig. 10(a) presents a comparison of the BER performance between MU-MDM and SU-MDM with the different loss functions, i.e., the L2-norm vs. the proposed one. As only EDs among the different data sequences (signal points) are considered, BER performance of L2-norm is optimized only in a ED sense but not in a bit-to-symbol mapping sense. It is clearly shown here that that the proposed loss function achieves 0.3 dB and 0.4 dB gain at a target BER of $10^{-3}$, as compared to the L2-norm loss function, for both SU-MDM and MU-MDM, respectively. In Fig. 10(b), furthermore, we investigate how effective the proposed loss function works for MU-MDM with the different levels of PNL. Even if we observe an obvious gain with the proposed loss function, its gain decreases with Level-3 PNL, i.e., 0.7 dB to 0.4 dB. It implies that the smaller the minimum ED between adjacent CWs, the more gain of increasing the HD between them.   

Fig. 11 presents a comparison of the BER performance for different levels of PNL with sparse and dense resource mapping for the proposed MU-MDM. As far as power-imbalance CB is considered with Level-3 PNL, however, dense resource mapping will be a baseline structure that includes sparse resource mapping as a special case. Whenever necessary, then, an option of Level-3 PNL along with dense resource mapping would set the constellation power for specific resources to zero, flexibly making the resource sparse as shown in Fig. 8. That is why the performance of dense resource mapping would be always the same or better than that of sparse resource mapping subject to the given level of PNL. In Fig. 11, we observe that Level-2 PNL outperforms Level-1 by 1 dB at a target BER of $10^{-3}$. Similarly, Level-3 outperforms Level-2 by 0.7 dB, i.e., a combined gain of 1.7 dB over Level-1, for sparse MU-MDM. Similarly, 0.8 dB performance gain is observed for dense MU-MDM. In fact, the BER performance gap between the dense and sparse CD-NOMA decreases with an increase in the level of PNL, showing only 0.1 dB gap between them. In fact, sparse CD-NOMA may be still comparably acceptable as long as Level-3 PNL is employed under the proposed architecture. In Fig. 11, we can also note that the BER performance gap between the dense and sparse CD-NOMA decreases with a higher level of PNL. Interestingly, in Level-3 PNL, the performance gap between sparse and dense CD-NOMA nearly disappears. This is attributed to a less stringent power constraint in Level-3 PNL, which leads to more room for optimizing the BER performance. 

Fig. 12 presents a comparison of the BER performances between the proposed designs for SU-MDM and MU-MDM. Our objective herein is to demonstrate that a performance of the proposed design for MU-MDM approaches that for SU-MDM, which serves as a lower-bound. Resorting to the conclusion from Fig. 10, we just consider the best-performing design, which is a dense CD-NOMA with Level-3 PNL, to be compared with SU-MDM. As shown therein, we observe that its performance approaches that of SU-MDM, with approximately 0.3 dB only, supporting our argument that the proposed AE architecture and the underlying training framework for CD-NOMA, without knowing other-user symbols, approaches the performance of SU-MDM with the same spectral efficiency. Meanwhile, we would like to highlight that the \textbf{Constraint A} discussed in the Section I is addressed in an implicit manner for the AE-based MU-MDM design as the AE sees all possible combination of symbols from all users in the course of designing the CBs. In particular, as the AE is trained for all the combination of the symbols, i.e., $M^{J}=4096$ combinations, the CB design of each user is implicitly interdependent via back propagation of the loss. It is to be noted that SU-MDM can be considered as MU-MDM design with explicit dependence (among users) in the symbols-to-CW mapping. However, in our MU-MDM case as the interdependency is not explicit as compared to SU-MDM, we were only able to approach the SU-MDM performance with 0.3 dB gap at $10^{-3}$ BER rather than perfectly achieving it. 

Furthermore, it is also compared with 8-QAM, which has the same bandwidth efficiency, i.e., 3bps/Hz, as 4-ary CB ($M$ = 4) over four orthogonal resources ($K$ = 4), overloaded with six users ($J$ = 6). The proposed design outperforms 8-QAM, e.g., by 1.5 dB and 2.1 dB for non-rectangular 8-QAM and rectangular 8-QAM, respectively. It can be translated into the maximum possible overloading gain that can be achieved by CD-NOMA as compared to OMA. Furthermore, in order to verify the current results, we also present a BER curve for QPSK, which is referred to as “Single-user BER bound” in Fig. 12 It corresponds to the BER of the same modulation scheme without overloading, serving as an ultimate lower bound on all possible designs. It is clearly shown that a performance of the proposed MU-MDM is now bounded by QPSK performance with a gap of approximately 0.74 dB at a BER of $10^{-3}$.

Fig. 13 presents a comparison of BER performance between the proposed DL-based design and conventional designs, which have a common feature of sparse resource mapping. The conventional designs include the conventional SCMA [10], [13] and DL-based SCMA [17]. First of all, we note that a performance of [17] is mainly limited by non-optimized bit-to-symbol mapping (associated with a loss function of L2-norm) and stringent power constraint (leve-1 PNL). However, it outperforms the power-balanced CB design in [13], simply because multi-user constellation can be further optimized with the AE-based design. Meanwhile, we note that another conventional design of SCMA in [13] follows a design discipline of allocating the different power for each CW, similar to our Level-3 PNL. It is interesting to note that such a power-imbalanced SCMA outperforms [17], even if its CB design relies just on a heuristic. It implies how effective to allows for the different power among the CWs. We finally note that the proposed design with Level-3 PNL outperforms the power-imbalanced CB design in [13]. As discussed in Section IV-C, this is attributed to the joint optimization capability of the DL-based design.

\begin{figure}[t]
\centering
\includegraphics[width=3.1in,height=2.6in]{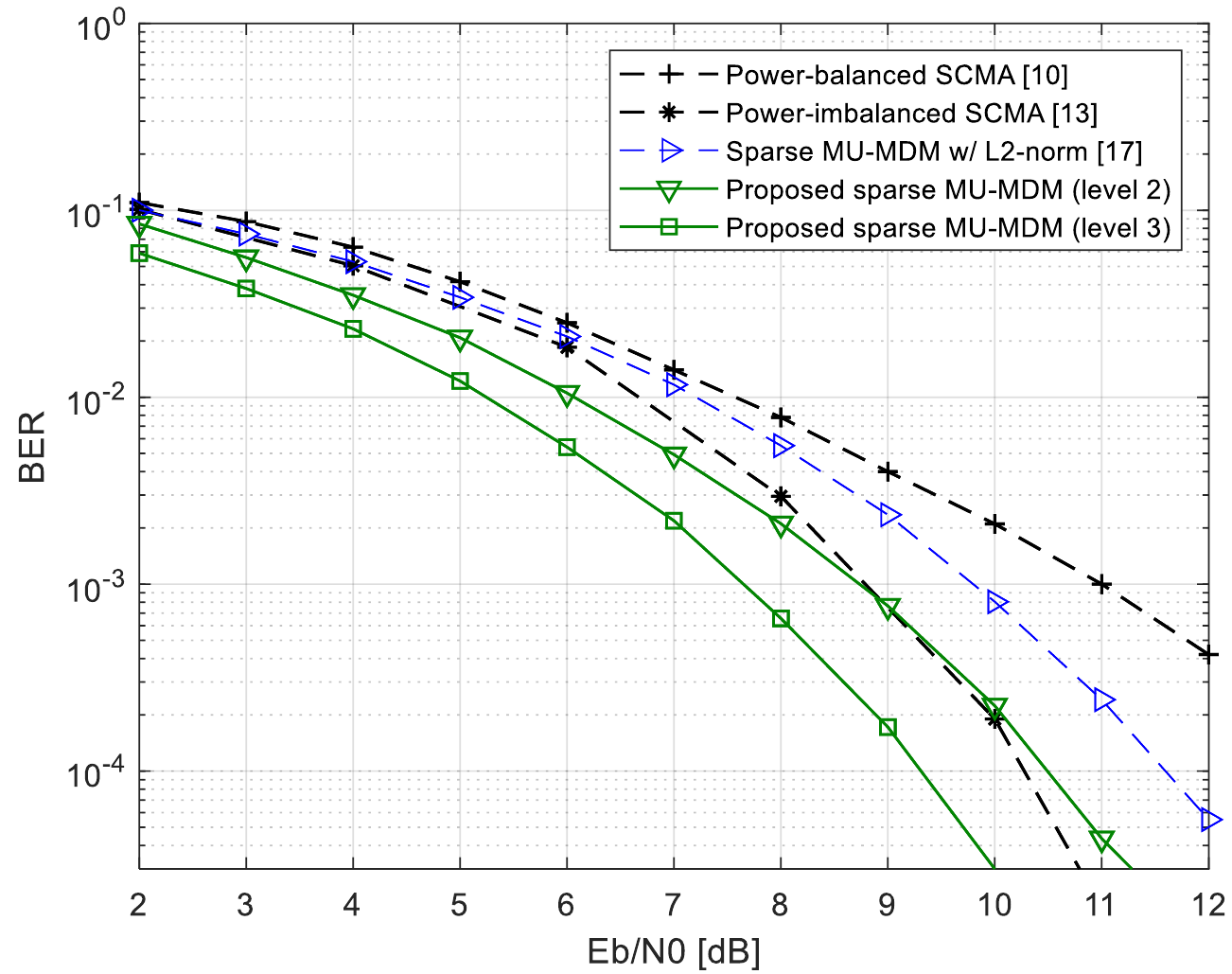}
  \captionsetup{font=small,justification=raggedright, singlelinecheck=false}
\caption{Comparison of BER performance: The proposed design vs. existing designs}
\label{fig_sim}
\end{figure}

\section{Conclusion}
This paper proposed a new deep learning-architecture for autoencoder to design a CD-NOMA scheme that aims at joint optimization of resource mapping and codebook with bit-to-symbol mapping. The proposed deep learning-based autoencoder learns a multi-user constellation, equivalently a codebook for CD-NOMA. We also presented the trainable architectures of deep learning-based designs for both SU-MDM and MU-MDM, which can be equivalently compared to each other. Based on them, our implementation demonstrated that a BER performance of our proposed architecture can approach single-user performance, which serves as a lower BER bound on the performance of deep learning-based design. The current achievement is mainly attributed to power-imbalanced codebook design and flexible resource mapping, along with a hyperparameterized loss function and corresponding training procedure to minimize the BER.

We note that both the existing and the proposed multi-user CBs are designed by assuming an AWGN channel. However, in the case of a fading channel, the inclusion of either conventional or NN-based equalizers at the receiver may transform the fading channels to an AWGN channel. Particularly for the uplink, as the user experiences different channels, a careful design of various channel compensation methods, such as close-loop/open-loop power control, should be an essential aspect of the proposed MU-MDM scheme in order to achieve consistent performances, as in the AWGN channel. Therefore, an interesting future research goal would be to investigate an NN-based multi-user channel equalizer or power control schemes that can be matched by the proposed scheme.

\appendices
\section{Proof of Proposition \ref{theorem1}}
Let $\varepsilon (K,M)$ denote an SER, which is defined as
\begin{align}\tag{A-1}
\varepsilon (K,M)=p[g(\mathbf{y})\ne i|f(i)],
\end{align}
for an equally likely symbol, indexed by $i=1,\cdots ,M$. In conventional communication theory, the encoder and decoder are designed to minimize the SER by solving the following optimization problem:
\begin{align}\tag{A-2}
({{f}^{*}},{{g}^{*}})=\underset{f,g}{\mathop{\arg \min }}\,\varepsilon (K,M).
\end{align}
However, in AE-based optimization, (4) is solved by replacing the loss function with the L2-norm function, which is given as
\begin{align}\tag{A-3}
L(\mathbf{r},\mathbf{\hat{r}})=L(\mathbf{r},g(f(\mathbf{r})))={{\left\| \mathbf{r}-g(f(\mathbf{r})) \right\|}_{2}}.
\end{align}
for an input vector $\mathbf{r}$ and its estimated $\mathbf{\hat{r}=}g(f(\mathbf{r}))$ with $\dim(\mathbf{r})=dim(\mathbf{\hat{r}})\ge {{\log }_{2}}M$. Furthermore, the encoder–decoder pair from the AE that was optimized by L2-norm in (A-3) is given as:
\begin{align}\tag{A-4}
(f_{AE}^{{{L}_{2}}},g_{AE}^{{{L}_{2}}})=\underset{(f,g)}{\mathop{\arg \min }}\,{{L}_{2}}\left( \mathbf{r},\mathbf{\hat{r}}\left| {{\theta }^{(f)}},{{\theta }^{(g)}} \right. \right).
\end{align} 

It is demonstrated in [27] that MLD minimizes SER, which is determined by the $K$, $M$, and SNR levels. Furthermore, as all $M$ symbols are transmitted with an equal likelihood, and AWGN is assumed; therefore ML-based detection and the minimum ED (MED)-based detection achieve the same SERs [29]. Therefore, 
\begin{align}\tag{A-5}
{{\varepsilon }_{MED}}(K,M)={{\varepsilon }_{ML}}(K,M)=\underset{f,g}{\mathop{\min }}\,\varepsilon (K,M),
\end{align}
where ${{\varepsilon }_{MED}}(K,M)$ denotes the SER with the minimum ED detection. Furthermore, in conventional communication theory, the optimal encoder–decoder pair that achieves ${{\varepsilon }_{MED}}(K,M)$ can be designed by solving following the optimization problem:
\begin{align}\tag{A-6}
& (f_{MED}^{*},g_{MED}^{*}) = \nonumber\\
 & \underset{(f,g)}{\mathop{arg\min }}\,p\left[ {{\left\| {{\mathbf{r}}_{i}}-g(f({{\mathbf{r}}_{i}})) \right\|}_{2}}>{{\left\| {{\mathbf{r}}_{j}}-g(f({{\mathbf{r}}_{i}})) \right\|}_{2}}|f({{\mathbf{r}}_{i}}) \right],\forall i,j\ne i, \nonumber
\end{align}
for the $k$-th symbol ${{\mathbf{r}}_{k}}$. 

If the AE is perfectly trained with the L2-norm, (A-4) for  the L2-norm loss function in (A-3) results in
\begin{align}\tag{A-7}
{{\left\| {{\mathbf{r}}_{i}}-g_{AE}^{{{L}_{2}}}(f_{AE}^{{{L}_{2}}}({{\mathbf{r}}_{i}})) \right\|}_{2}}\to 0.
\end{align}
It should be noted that the optimal encoder-decoder pair for minimum ED detection in (A-6) can be achieved when (A-7) goes to zero. Therefore, the optimal encoder-decoder pair in (A-6) can be achieved by globally minimize (A-4). Proposition 1 is thus established through the following conclusion: 
\begin{align}\tag{A-8}
(f_{AE}^{{{L}_{2}}},g_{AE}^{{{L}_{2}}})=(f_{MED}^{*},g_{MED}^{*}).
\end{align}
\begin{proof}[\killproofname] 
\end{proof}


%


\ifCLASSOPTIONcaptionsoff
  \newpage
\fi

\end{document}